\documentclass[prd,superscriptaddress,showpacs,onecolumn]{revtex4-2}
\usepackage{bm}
\usepackage{amsfonts}
\usepackage{latexsym}
\usepackage[latin1]{inputenc}
\usepackage{graphicx}
\usepackage{amsmath}
\usepackage{palatino}
\usepackage{mathpazo}
\usepackage{textcomp}
\usepackage{float}
\usepackage{booktabs}
\usepackage{dcolumn}
\usepackage{hyperref}
\usepackage{amsmath}
\usepackage{xcolor}
\usepackage{orcidlink}
\usepackage[caption=false]{subfig}
\usepackage{commath}

\hypersetup{colorlinks,citecolor=blue}
\def\jnl@style{\it}
\def\aaref@jnl#1{{\jnl@style#1}}
\def\aaref@jnl#1{{\jnl@style#1}}
\def\aj{\aaref@jnl{AJ}}
\def\apj{\aaref@jnl{ApJ}}
\def\apjl{\aaref@jnl{ApJ}}
\def\apjs{\aaref@jnl{ApJS}}
\def\apss{\aaref@jnl{Ap\&SS}}
\def\aap{\aaref@jnl{A\&A}}
\def\aapr{\aaref@jnl{A\&A~Rev.}}
\def\aaps{\aaref@jnl{A\&AS}}
\def\mnras{\aaref@jnl{Mon.~Not.~Roy.~Astron.~Soc.}}
\def\prd{\aaref@jnl{Phys.~Rev.~D}}
\def\prc{\aaref@jnl{Phys.~Rev.~C}}
\def\prl{\aaref@jnl{Phys.~Rev.~Lett.}}
\def\qjras{\aaref@jnl{QJRAS}}
\def\skytel{\aaref@jnl{S\&T}}
\def\ssr{\aaref@jnl{Space~Sci.~Rev.}}
\def\zap{\aaref@jnl{ZAp}}
\def\nat{\aaref@jnl{Nature}}
\def\aplett{\aaref@jnl{Astrophys.~Lett.}}
\def\apspr{\aaref@jnl{Astrophys.~Space~Phys.~Res.}}
\def\physrep{\aaref@jnl{Phys.~Rep.}}
\def\physscr{\aaref@jnl{Phys.~Scr}}
\def\commat{\aaref@jnl{Comm.~Math.~Phys.}}
\def\science{\aaref@jnl{Science}}
\def\cqg{\aaref@jnl{Classical Quant.~Grav.}}
\def\jpcs{\aaref@jnl{JPCS}}
\def\ijmpd{\aaref@jnl{Int.~J.~Mod.~Phys.~D}}
\def\grg{\aaref@jnl{Gen.~Relat.~Gravit.}}
\def\rpp{\aaref@jnl{Rep.~Prog.~Phys.}}
\def\npa{\aaref@jnl{Nucl.~Phys.~A}}
\def\lrr{\aaref@jnl{Living Rev.~Rel.}}
\def\jcap{\aaref@jnl{J.~Cosmology Astropart.~Phys.}}
\def\rmp{\aaref@jnl{Rev.~Mod.~Phys.}}
\def\epjc{\aaref@jnl{Eur.~Phys.~J.~C}}

\allowdisplaybreaks[1]

\begin{document}

\title{\textbf{A model of dark matter-dark energy interaction with some cosmic consequences}}

\author{J.K. Singh}
\email{jksingh@nsut.ac.in}
\affiliation{Department of Mathematics, Netaji Subhas University of Technology, New Delhi, India}
\author{Ritika Nagpal}
\email{ritikanagpal.math@gmail.com}
\affiliation{Department of Mathematics, Vivekananda College, University of Delhi, New Delhi, India.}
\date{\today }


\begin{abstract}

In this study, we will look at an interacting dark energy model. In the framework of Friedmann-Robertson-Walker (FRW) space-time, we have made the hypothesis of an interacting scheme between two fields (dark matter (DM) and dark energy (DE)). The evolution of the dark energy model in the above-mentioned spatially and homogeneous space-time has been studied from the viewpoint of interaction between DM and DE. This interacting scenario has been established by choosing an appropriate ansatz of the scale factor $a(t)$. By computing cosmological parameters (geometrical and physical) such as deceleration parameter $q$, energy density $\rho$, pressure $p$, equation of state parameter $\omega$, and density parameter $\Omega$, we have performed energy conditions, scalar field description, and refined swampland conjecture to support the interaction model that we have developed.

\end{abstract}

\maketitle

\color{black} 


\section{Introduction}

\label{sec1}

\qquad One of the biggest mysteries in cosmology today is the physical explanation for the late-time cosmic speed-up. The perplexing nature of dark energy (DE) and dark matter (DM) has been a focus of theoretical physics and cosmology for decades. The multiple statistical observational data sets have explained this physical mechanism including \cite{Reiss1998}, \cite{Perlmutter1999}, \cite{Tegamark2004}, \cite{SELJAK2004}, \cite{BAO2006}, \cite{SDSS2006}, \cite{CMB2007}, \cite{Hinshaw2013}. Some tenable ideas for the hypothetical form of DE include an infra-red adjustment to Einstein's general theory of relativity (for some reviews, see \cite{ejcope}, \cite{snoj}). Therefore, the evolution of the modern is governed by DE, which makes up around 68$\%$ of the total energy density. According to several DE models, the expansion of the Universe is influenced by an extra, unidentified field. \newline

\qquad The most simplest and straightforward theory of DE is the one based on the cosmological constant $\Lambda$, whose equation of state $(EoS)$ parameter $\omega_ \Lambda=\frac{p_\Lambda}{\rho_\Lambda}=-1$. The $\Lambda$CDM model agrees well with the most recent cosmological observations, such as those from the experiment \cite{parplanck}, and its parameters have been precisely determined using the most recent observational data. The $\Lambda$CDM model, on the other hand, has consistently been cursed by many open theoretical problems like fine-tuning and cosmic-coincidence concerns \cite{vsahni}. Due to this, several extensions \cite{xzhan1}, \cite {msas}, \cite{rmurg}, \cite{lfeng}, \cite{apou} to the fundamental $\Lambda$CDM cosmology have drawn a lot of interest, including the capacity for vacuum energy to interact with $\Lambda$CDM. Weinberg (1989) \cite{Weinberg1989} advised modifying the gravity and moving toward alternating theories of gravity with an improvement in the goodness of fit, such as \cite{parplanck1} and \cite{nagh} due to the large discrepancy between theoretical and observational predictions on the value of $\Lambda$ and to ease the tension between the $\Lambda$CDM and the observational data. The slow roll scalar field has been widely covered in the literature as an alternative for the $\Lambda$CDM model \cite{evlin}. The significance of adopting the DE candidate in the form of a scalar field comprehending the integration of early and late time cosmic acceleration has consequently received strong acceptance\cite{ejcope}.\newline

\qquad Besides the flaws in the standard cosmological model, many experimental results remain unknown. In certain research, the $\Lambda$CDM model has been altered by coupling the $H_{2.34}$ data with the DDE model (dynamical dark energy model). A variety of models allow the evolution of DDE models based on the observations $BAO$, $CMBR$, and $SNIe$ data and it can be studied in the work discussed in \cite{eaub} and \cite{gbzh}. The study in this model further explains how to match DDE models with $H_{2.34}$ data. The measured value of the Hubble parameter $H$ at redshift $z=2.34$ is $222 \pm 7 kms^{-1} Mpc^{-1}$ mentioned in many observations of $BOSS$ data \cite{tdel}. It has been observed that this value is less than the value predicted by the $\Lambda$CDM model. In research work, \cite{agva} and \cite{rygu} some further efforts are used to show the comparison between the theoretical and observed value. \newline

\qquad This study aims to develop a cosmological model that takes into account unique phenomenological types of interaction between DM and DE. We employ the model-independent approach that involve some parametrization schemes that reduces the tension between the observational datasets \cite{egmf}, \cite{edva} with standard $\Lambda$CDM model. This idea perfectly describes the late acceleration of the Universe. Therefore in this work, we shall reveal new features for the interaction of both components DM and DE. \newline

\qquad Numerous strategies have been designed to uncover the physical characteristics of DM and DE because the nature of the dark sector is not completely understood. Many techniques have been devised to reveal the physical characteristics of DM and DE because of a lack of understanding of the dark sector. We are still researching theories that could offer an alternative method for analyzing the characteristics of the dark sector and ideally enable us to discriminate among the various theoretical theories. According to certain hypotheses, DM and DE could interact in this way to disclose new aspects \cite{bwang}, \cite{mavan1}, \cite{mavan2}. Numerous interactions such as non-minimally coupled theories \cite{tkoi}, \cite{lamen} where the entire Lagrangian contains a particular interaction term has been presented. \newline

\qquad A possible explanation for the interaction of DM and DE holds that any two distinct matter fields can interact with each other either in particle physics or on a more theoretical level. The cosmology community is enthusiastic about such a particular phenomenological theory because of its potential for a diverse range of outputs. Understanding the dark sector of the Universe is the big fuss nowadays and it's equivalent to understanding the interacting models of DM and DE that have been extensively researched. Also, it has been simulated that interacting models are the viable solution of the most prevailing issues of the standard $\Lambda$CDM model (\textit{i.e.} coincidence and cosmological constant problem) as in these model DE decompose into DM \cite{wzim}, \cite{femc}. Some research to alleviate or resolve the paradox of the coincidence problem makes reasonable assumptions regarding a DM-DE relationship. Because interaction between these dark matter and dark energy is possible in field theory \cite{smic}, we do not comprehend the nature of DM and DE. Seeing the energy density of DM and DE dynamically in this fashion lower the coincidence issue. Numerous studies that show the connection between DM and DE interaction have already been reported in this respect, including those by, \cite{smic}, \cite{eabda1}, \cite{keshav}, \cite{vsal}. See \cite{bwang2} for a more thorough list of references on interaction evidence that has recently been found as well as express the discussion of theoretical and cosmological characteristics. \newline

\qquad Learning-interaction theory has the benefit of enabling one to transform the DE equation of state parameter (EoS) from quintessence to phantom, indicating that the effective quintom nature of the DE EoS parameter is applied. Furthermore, it has been found that the interaction theory is especially good at bridging the Hubble constant $(H_0)$ gap between global and local measurements.\newline

\qquad As a result, a variety of linear and non-linear functions are described in the literature, and one can supposedly introduce some new functions because there is no such overarching criterion for employing interaction functions. It is typically cumbersome to discuss the dynamics of an interacting model utilizing a non-linear interaction function. Studies like \cite{lpchi}, \cite{fare}, and \cite{wyang} are not frequently found in the literature for having the choice of interacting functions. However, it is always fascinating to investigate the dynamics of the Universe using non-linear interacting models to see if we can discover anything new from them. Due to the encouragement provided by the interacting situations, we use a model-independent methodology in this investigation to assess an interacting model.\newline

\qquad The work of the paper has been arranged in the manner described below: A basic overview of general relativity and the most pressing cosmological issues are given in Section 1. The fundamental setup of the governing equations of matter-energy interaction within GR, examining the spacetime metric, and formulating the Einstein Field equations in GR is described in Section 2. In the next Section 3, we discussed the outcomes of the governing equations acquired in Section 2. We used a straightforward parametrization method in Section 4 to arrive at the solutions to the field equations and discussed the model. Using various external datasets, we limited the model parameters in Section 4. The next sections discuss some in-depth analyses of the model using graphical representations of how physical parameters behave. We explored the energy conditions of the model in Section 5. The viability of our model using the velocity of sound is explored in Section 6 and examined the Swampland conjecture in Section 7. We draw a conclusion on our findings in Section 8.\newline

\section{Governing equations of matter-energy interaction }

\label{sec2}
 
We begin with the FRW metric which is spatially homogenous and isotropic provided in the form of spherically symmetric geometry, 
\begin{equation}
ds^{2}=-dt^{2}+a(t)^{2}\left[ \frac{dr^{2}}{1-\kappa r^{2}}+r^{2}(d\theta^{2}+\sin ^{2}\theta d\phi ^{2})\right] .  \label{1}
\end{equation}
As a function of cosmic time $t$, the function $a(t)$ represents the scale factor of the Universe in this context. The term $\kappa$ denotes the curvature constant, which takes the values $0$, $-1$, or $1$ to represent flat, open, or closed geometry respectively.\newline

For the FRW spherically symmetric metric in GR, let us present the goverening equations called as Einstein field equations (EFEs) of the form 
\begin{equation}
R_{\alpha \beta }-\frac{1}{2}Rg_{\alpha \beta }={8 \pi G}T_{{\alpha \beta }},  \label{2}
\end{equation}%
where the energy momentum tensors of the various components of the Universe such as radiation, baryons, DM, and DE are represented in the right hand side of the above equation while the geometry of the Universe is denoted in the left hand side. From the work presented in \cite{keshav}, where the field equations involving all the densities in one equation, we have utilized the field equations as a foundational framework to delve into the realm of interacting scenarios within our model,
\begin{equation}
\Big(\frac{\dot{a}}{a}
\Big)^{2}+\frac{k}{a^{2}} =\frac{8\pi G}{3}[\rho _{\textit{Radiation}}+\rho _{\textit{Baryons}}+\rho _{\textit{Dark matter}}+\rho
_{\textit{Dark energy}}]=\frac{8\pi G}{3} \rho _{\textit{Total}}\text{,}  \label{3}
\end{equation}%

\begin{equation}
 2\frac{\ddot{a}}{a}+\Big(\frac{\dot{a}}{a}
\Big)^{2}+\frac{k}{a^{2}}
=-8\pi G[p_{\textit{Radiation}}+p_{\textit{Baryons}}+p_{\textit{Dark matter}}+p_{\textit{Dark energy}}]=-8\pi G p_{\textit{Total}}\text{.}  \label{4}
\end{equation}
In this case, the time derivative is represented by an overhead dot and the Hubble parameter $\frac{\dot a}{a}$, is shown by the symbol $H$. The energy densities of various components of the Universe are represented by the symbols $\rho _{\textit{Radiation}},\rho _{\textit{Baryons}},\rho _{\textit{Dark matter}},\rho_{\textit{Dark energy}}$, whereas the pressures of these components are represented by $p_{\textit{Radiation}}\\ p_{\textit{Baryons}}, p_{\textit{Dark matter}}, p_{\textit{Dark energy}}$. From this point on, we will refer to the various energy densities and pressures in using the following notations.

\begin{enumerate}
\item $\rho_{\textit{Radiation}}$ and $p_{\textit{Radiation}} $ as $\rho_{R}$ and $p_{R}$
\item $\rho_{\textit{Baryons}}$ and $p_{\textit{Baryons}}$ as $\rho_{B}$ and $p_{B}$
\item $\rho_{\textit{Dark matter}}$ and $p_{\textit{Dark matter}}$ as $\rho_{DM}$ and $p_{DM}$
\item $\rho_{\textit{Dark energy}}$ and  $p_{\textit{Dark energy}}$ as $\rho_{DE}$ and $p_{DE}$
\end{enumerate}

The Bianchi identity states that $G_{\alpha \beta}^{;\beta}=0$ results in $T_{\alpha \beta}^{;\beta}=0$. The resulting continuity equation can be seen below. \newline
\begin{equation}
\dot{\rho}_{Total}+3\frac{\dot{a}}{a} (\rho _{Total}+p_{Total})=0\text{.}  \label{5}
\end{equation}

In cosmology, particularly in theories of DE, the interaction term between the Hubble parameter and the energy density of matter is frequently used. These are some potential justifications for choosing this term. 
\begin{itemize}
\item The interaction term could have the potential to explain the coincidence problem. The interaction of dark energy with matter, which causes a co-evolution of their energy densities, can be used to explain the coincidence problem that has already been mention in the introduction. Since the Hubble parameter determines the expansion rate of the universe, therefore, including the Hubble parameter $H$ in the interaction term enables a connection to be made between the evolution of energy densities and the rate of expansion of the universe.
\item The Hubble parameter generally changes in response to changes in the matter and energy content of the universe, according to the standard cosmological model. Up to now, this expansion has been influenced by the presence of an interaction term such as $\rho_{DM}$, leading to intriguing cosmic dynamics. For example, in some models with an interaction term, the equation of state for dark energy can change, which may help to explain observations of the accelerated expansion of the Universe. 
\item Studies suggest a cosmological model with an interaction term that could modify the dark energy equation of state. It has been noted that the interaction term significantly affects the evolution of dark energy and the history of the universe's expansion, leading to a time-varying equation of state for dark energy. The work clarified the characteristics of dark energy and showed that the model agrees with the Planck data. Here are a few sources in this direction \cite {harkoasadi},\cite{weqiang}, \cite{andrea}. There are various observable evidences that have been offered to describe the interaction between dark matter and dark energy, which is a topic of ongoing study in cosmology. Here are few sources that examine various facets of this subject \cite{wzimdahl}, \cite{elcioa}, \cite{bwange}, \cite{christian}, \cite{ferlito}, \cite{valentino1}, \cite{valentino2}, \cite{garia}, \cite{noon}, \cite{noon1}, \cite{bernui}, \cite{vagno1} and \cite{vagno2}.
\end{itemize}

There is currently no convincing observable evidence that the interaction term is connected to energy density of dark energy $\rho_{DE}$ rather than dark matter $\rho_{DM}$.  Nonetheless, there are several theoretical arguments and observational indicators that lean towards supporting this concept. Since dark energy has a higher energy density than dark matter, it is important to keep in mind that the effects of dark energy on the cosmos usually become apparent only after the universe has developed into a well-organized state. On the other hand, dark matter continues to play a significant role in the development of galaxies and other largescale structures. This interaction term may well have significant effects on the evolution of the universe, influencing the expansion rate, the formation of structure, and the production of galaxies. One possible result of the interaction term proportional to dark matter energy density $\rho_{DM}$ is that it might offer a different explanation for the observed acceleration of the universe expansion. Instead of relying on a cosmic constant (\textit{i.e.} a constant dark energy density), the acceleration might be driven on by an interaction between dark matter and dark energy. Through studies of the cosmic microwave background radiation, large-scale structure, and the evolution of structure, the kind of the interaction term can be deduced \cite{shuoc}, \cite{yunliu}.  

Let us think about the interacting scenario in this article after the motivation talk mentioned above and in the introductory section. The equations follow when we suppose that DM and DE are the two most dominant components in gravitational interactions. 
\begin{equation}
\dot{\rho}_{R}+3\frac{\dot{a}}{a}
(\rho _{R}+p_{R})=0\text{,}  \label{6}
\end{equation}%
\begin{equation}
\dot{\rho}_{B}+3\frac{\dot{a}}{a}
(\rho _{B}+p_{B})=0\text{,}  \label{7}
\end{equation}%
\begin{equation}
\dot{\rho}_{DM}+3\frac{\dot{a}}{a}
(\rho _{DM}+p_{DM})=Q(t)\text{,}  \label{8}
\end{equation}%
and 
\begin{equation}
\dot{\rho}_{DE}+3\frac{\dot{a}}{a}
(\rho _{DE}+p_{DE})=-Q(t)\text{,}  \label{9}
\end{equation}%
where $Q(t)$ denotes the rate of energy transfer between the two dark sectors in these equations, with $Q(t)>0$ signifying a transfer of energy from the DM to DE sector and $Q(t)<0$ signifying the inverse. However, we are still unsure of its precise nature. Certain plausible forms of $Q$ should be presumed in order to study the subject of dark sector interaction. In this article, we explore the form
\begin{equation}
Q=3\epsilon \frac{\dot{a}}{a}
\rho _{DM},  \label{10}
\end{equation}%
where the coupling constant is $\epsilon $. Since the interaction term must be proportional to the inverse unit of time \cite{amendola}, \cite{pavon1}, \cite{guo} for the preceding equations (\ref{8}) and (\ref{9}) to hold the continuity law, the interaction term must be exactly proportional to the Hubble parameter $H$. As a result, to reflect this feature, we have chosen the form of $Q=3\epsilon \frac{\dot{a}}{a} \rho_{DM}$.

\section{Outcome of the governing equations }
Let us say that $\omega=\frac{p}{\rho}$ represents the equation of state parameter for different Universe components. The answer is then obtained from the field equations (\ref{6}), (\ref{7}), (\ref{8}) and (\ref{9}) by combining them (\ref{10}),
\begin{equation}
\rho _{R}=\frac{A_1}{a^{4}}\text{,}  \label{10a}
\end{equation}%
\begin{equation}
\rho _{B}=\frac{A_2}{a^{3}}\text{,}  \label{10b}
\end{equation}%
\begin{equation}
\rho _{DM}=\frac{A_3}{a^{3(1-\epsilon)}}\text{,}  \label{11}
\end{equation}%
and%
\begin{equation}
\dot{\rho}_{DE}+3\frac{\dot{a}}{a}\left( 1+\omega _{DE}\right) \rho _{DE}=-3H\epsilon A_3 a^{-3(1-\epsilon )}\text{,}  \label{12}
\end{equation}%
where $A_1$, $A_2$ and $A_3$ are integrating constants. The redshift $1+z=\frac{a_{0}}{a}$ will now be defined. Moving forward, we will be using the conventional form and normalized the scale factor $a_{0}=1$. So we can now write equations (\ref {10a}), (\ref {10b}) and (\ref{11}) as,

\begin{equation}
\rho _{R}=A_1\left( 1+z\right) ^{4}\text{, }\rho _{B}=A_2\left(1+z\right) ^{3}\text{, }\rho _{DM}=A_3\left( 1+z\right) ^{3-3\epsilon }\text{,}  \label{12a}
\end{equation}%
Additionally, the equations for $\rho _{DE}$ and $p_{DE}$ can be constructed using the following geometrical parameters:
\begin{equation}
p_{DE}=\frac{-1}{8\pi G}\left[ \left( 2q-1\right) H^{2}-k\left( 1+z\right) ^{2}%
\right] -\frac{1}{3}A_{1}\left( 1+z\right) ^{4}\text{,}  \label{13}
\end{equation}

\begin{equation}
\rho _{DE}=\frac{3}{8\pi G}\left[ H^{2}+k\left( 1+z\right) ^{2}\right] -\left[
A_{1}\left( 1+z\right) ^{4}+A_{2}\left( 1+z\right) ^{3}+A_{3}\left(
1+z\right) ^{3-3\epsilon }\right] \text{.}  \label{14}
\end{equation}
Now let us define the density parameter ($\Omega $) as $\Omega_j=\frac{\rho_j}{\rho_{critical}}$, where $\rho_{critical}=\frac{3H^{2}}{8\pi G}$ denotes the critical density and the subscript $j=R$, $B$, $DM$ denotes Radiation, Baryons, and Dark matter respectively. Then, there is an expression as

\begin{equation}
\Omega _{R}=\Omega _{0,R}\left( 1+z\right) ^{4}\left( \frac{H_{0}}{H}\right)
^{2}\text{, }\Omega _{B}=\Omega _{0,B}\left( 1+z\right) ^{3}\left( \frac{H_{0}%
}{H}\right) ^{2}\text{, }\Omega _{DM}=\Omega _{0,DM}\left( 1+z\right)
^{3-3\epsilon }\left( \frac{H_{0}}{H}\right) ^{2}\text{,}  \label{14a}
\end{equation}%
where the subscript $0,j$ denotes the values of the multiple cosmological parameters and $\Omega_{0,j}$ denotes the density parameter of radiation, baryons and dark matter at the current moment ($t=t_{0}$ or $z=0$)  In terms of the density parameter, the FRW equation (\ref {3}) can be stated as
\begin{equation}
\Omega _{DE}=\left( 1+\Omega _{k}\right) -\left( \frac{H_{0}}{H}\right) ^{2}%
\left[ \Omega _{0,R}\left( 1+z\right) ^{4}+\Omega _{0,B}\left( 1+z\right)
^{3}+\Omega _{0,DM}\left( 1+z\right) ^{3-3\epsilon }\right] \text{,}  \label{14b}
\end{equation}%
taking into account $\frac{k\left( 1+z\right) ^{2}}{H^{2}}=\frac{k}{a^{2}H^{2}}=\Omega _{k}$. Lastly, we have 
\begin{equation}
p_{DE}=-\frac{1}{8\pi G}\left[ \left( 2q-1\right) H^{2}-k\left( 1+z\right) ^{2}%
-\Omega _{0,R} H_{0}^{2}\left( 1+z\right) ^{4}\right] \text{,}  \label{14c}
\end{equation}

\begin{equation}
\rho _{DE}=\frac{3}{8\pi G}\left[ H^{2}+k\left( 1+z\right) ^{2}-H_{0}^{2}[\Omega _{0,R}\left( 1+z\right) ^{4}+\Omega _{0,B}\left( 1+z\right)
^{3}+\Omega _{0,DM}\left( 1+z\right) ^{3-3\epsilon }]\right] \text{,}  \label{14d}
\end{equation}%
and

\begin{equation}
\omega _{DE}=\frac{1}{3}\frac{\left( 2q-1\right) H^{2}-k\left( 1+z\right)
^{2}-H_{0}^{2}\Omega _{0,R}\left( 1+z\right) ^{4}}{H^{2}+k\left( 1+z\right)
^{2}-H_{0}^{2}\left[ \Omega _{0,R}\left( 1+z\right) ^{4}+\Omega _{0,B}\left(
1+z\right) ^{3}+\Omega _{0,DM}\left( 1+z\right) ^{3-3\epsilon }\right] }
\label{14e}
\end{equation}

It is now evident from the calculations above that we have a fully deterministic solution of the field equations for a known form of any geometrical parameter such as scale factor, Hubble parameter, or deceleration parameter which allows us to explain the features of the model. A form of the physical parameter is equally considered in the literature \cite{ritzepjc2} to discuss the solution of the field equation. A simple mathematical way to approach the this idea of solving the system of equations is the model-independent way that considers a parametrizatio scheme \cite{pacif1}, \cite{pacif2}. A few studies using a parametrization scheme to study the dark energy model can be found in the literature \cite{ritzepjc1}, \cite{ritzijgmmp}, \cite{ritzass}, \cite{ritzprd}, \cite{ritzannals}. Cosmic parametrization, a method independent of the model and a natural way to talk about cosmological dynamics, is a functional form of any geometrical or physical parameter. \\

The scale factor $a(t) = sinh(\beta t)$ was first derived in papers \cite{pradhan1}, \cite{pradhan2}, \cite{pradhan3} and \cite{pradhan4} and the scale factor  $a(t) = sinh(\beta t)$ is generalized in (2013) as $a(t) = [sinh(\beta t)]^{1/n}$ \cite{cha}, \cite{pradhan5}, \cite{pradhan6}, \cite{pradhan7} and \cite{pradhan8}. In line with this, we address this situation in our current work by adopting an appropriate parametrization of the scale factor $a(t)$ 
\begin{equation}
a(t)=[sinh(\beta t)]^{1/n},  \label{15}
\end{equation}%
where $n$ and $\beta$ are arbitrary constants that serve as the model's parameters and are constrained by a number of experimental datasets.  Here, $n$ is dimensionless and $\beta$ has dimension of $t^{-1}$. \\

The motivation for choosing $a(t)$ in Eq. (\ref{15}) has some mathematical significance. The presence of the $sinh$ function in the parameterization of scale factor $a(t)$  in cosmology is motivated by the desire to have a model that can describe both early-time inflationary expansion and late-time accelerated expansion of the universe, while also being consistent with observational data. The parametrization of scale factor has several desirable properties that make it a suitable choice for this purpose. First, a linear combination of an exponential function grows exponentially for large values of its argument. This behaviour is similar to the exponential expansion of the universe during the inflationary epoch. Second, the function also exhibits a more gradual exponential growth for smaller values of its argument, which makes it well suited to describe the late-time accelerated expansion of the universe that is observed today. Finally, this parametrization is a smooth and continuous function that can be easily differentiated, making it a useful tool for data analysing for evolution of the universe. In the given parametrization of $a(t)$, the model parameter $\beta$ is related to the rate of expansion of the universe, and model parameter $n$ determines the curvature of the universe. This parameterization is frequently employed in many studies because it enables the creation of a consistent model that can account for a wide range of cosmological situations, including as inflation, radiation dominance, matter dominance, and dark energy dominance. Overall, the parametrization is chosen as a scale factor because it offers a straightforward and adaptable model that can precisely represent the evolution of the cosmos from its origin to the present, while still being in line with experimental facts. When the densities of dark matter and dark energy are equal,\textit{i.e.,} in the intermediate epoch, the universe is in a stage when matter predominates over dark energy. The proportional contributions of these two components during this time govern the behavior of the universe, and the behavior of the scale factor is dependent on the particular cosmological model being taken into account.\\

In our study, for the choice of scale factor that we have taken into account, we have evaluated the behavior at intermediate times, when dark matter and dark energy are equal. It can be determined by setting $\rho_{DM}$ and $\rho_{DE}$ to be equal in the Friedmann equations or taking the ratio of density parameter of dark energy to dark matter equal to $1$ \textit{i.e.}, $\frac{\Omega_{DE}}{\Omega_{DM}}=1$. For model parameter $n= 1.1$, the ratio is approximately $1$ around $z = 1.65$, whereas for $n= 1.5$, the ratio is approximately $1$ around $z = 0.56$, which is closer to the estimated redshift as per standard $\Lambda CDM$ model (\textit{i.e.} approximately $z = 0.7$), which corresponds to a time of about $6$ billion years after the Big Bang. Therefore, we noticed the behavior of the scale factor at intermediate times depends on the specific values of the parameters in parameterization. Hence, choices of parameters, the scale factor may exhibit interesting features such as transient acceleration or deceleration, or oscillatory behaviour.\\

In addition, we have evaluated the value of Hubble parameter at redshift $z = 2.34$ for the considered parametrization. For $n=1.1, H_{2.34} \approx 160.021$, $n=1.5, H_{2.34}  \approx 233.675$, $n=2, H_{2.34}  \approx 363.539$. The value of $H_{2.34}$ at $z = 2.34$ is quite closer with the observation of BOSS data mentioned in the introduction.\\

Model parameters in the geometrical parameters like scale factor, the Hubble parameter or the deceleration parameter, are typically used to describe how the behaviour of the cosmos evolves through time. Contrarily, the interaction between dark matter and dark energy density parameters, which can affect the evolution of cosmos, describes the rate of energy transfer between these two components. According to the interaction theory, in models with a positive interaction parameter, the transfer of energy from dark energy to dark matter accelerates the rate at which the universe is expanding. Generally, depending on the particular cosmological scenario under consideration, the relationship between the model parameters of the scale factor $n$ and $\beta$ and the interaction parameter $\epsilon$ affects the behaviour of the universe and the interpretation of observational evidence.

The dynamics of the model with the considered parametrization will be covered in the following section.

\section{The Model}

\label{sec3}

The form of scale factor $a(t)$ taken into account in equation (\ref{15}) provides a smooth dynamics of the expanded Universe. The explicit expression of the Hubble parameter is easily produced by equation (\ref{15}) as,

\begin{equation}
H(t)=\frac{\beta coth(\beta t)}{n}\text{,}  \label{16}
\end{equation}
and 
\begin{equation}
q(t)=n(1-tanh^2(\beta t))-1\text{.}  \label{16a}
\end{equation}

Using the initial big bang scenario (at $t=0$, \textit{i.e.,} $a=0$) which sets the integration constant $c$ to zero. The equation for the $t-z$ relationship is as follows: 
\begin{equation}
t(z)=\frac{1}{\beta} sin h^{-1}\big(\frac{sin h(\beta t_0)}{(1+z)^n}\big) ,\label{16b}
\end{equation} 
therefore, in terms of redshift $z$, we can state the expressions of $H(z)$ and $q(z)$ as follows:

\begin{equation}
H(z)=H_0 (1+z)^n sec h(\beta) \sqrt{1+\frac{sin h(\beta)^2}{(1+z)^{2n}}},  \label{17}
\end{equation}%

\begin{equation}
q(z)=n \big[1-\frac{(sin h(\beta))^2}{(1+z)^{2n}+(sin h(\beta))^2}\big]-1.  \label{18}
\end{equation}

According to observational datasets, the current value of deceleration parameter is roughly $q \approx -0.55$. Therefore, using the equation (\ref{18}), we can establish a relationship between the model parameters $n$ and $\beta$ which yields some values listed below.\\
\begin{table}[ht]
\caption{ Values of $\beta$ for different choices of $n$}
\begin{center}

\begin{tabular}{l c c c r} 
\hline\hline
\\ 
$n$ & \,\,\,\,\, $0.5$ \,\,\,  &  \,\, \,$1$  \,\, \, &  \,\, \, $1.5$ \,\,\,   & \,\, \, \,\,\, $2$ \,\,\,\\ 
\\
\hline 
\\
$\beta$ & $0.32745$ & $0.95406$ & $1.2099$ & \,\,\,\,\,\, $1.37728$
\\
\\ 
\hline\hline  
\end{tabular}    
\end{center}
\end{table}

\begin{figure}[H] 
\begin{center}
$%
\begin{array}{c@{\hspace{.1in}}cc}
\includegraphics[width=3.5 in, height=2.5 in]{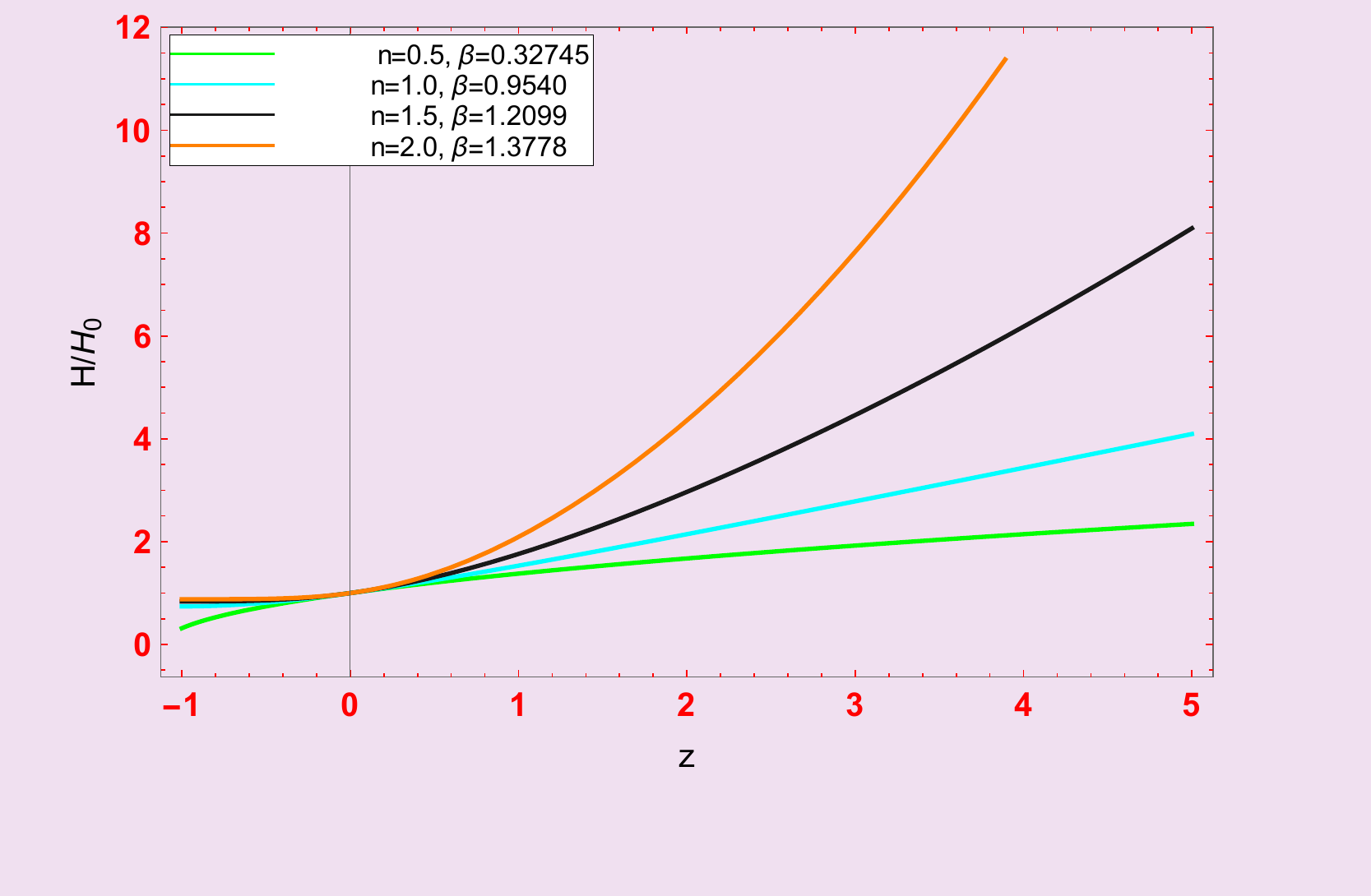} & %
\includegraphics[width=3.5 in, height=2.5 in]{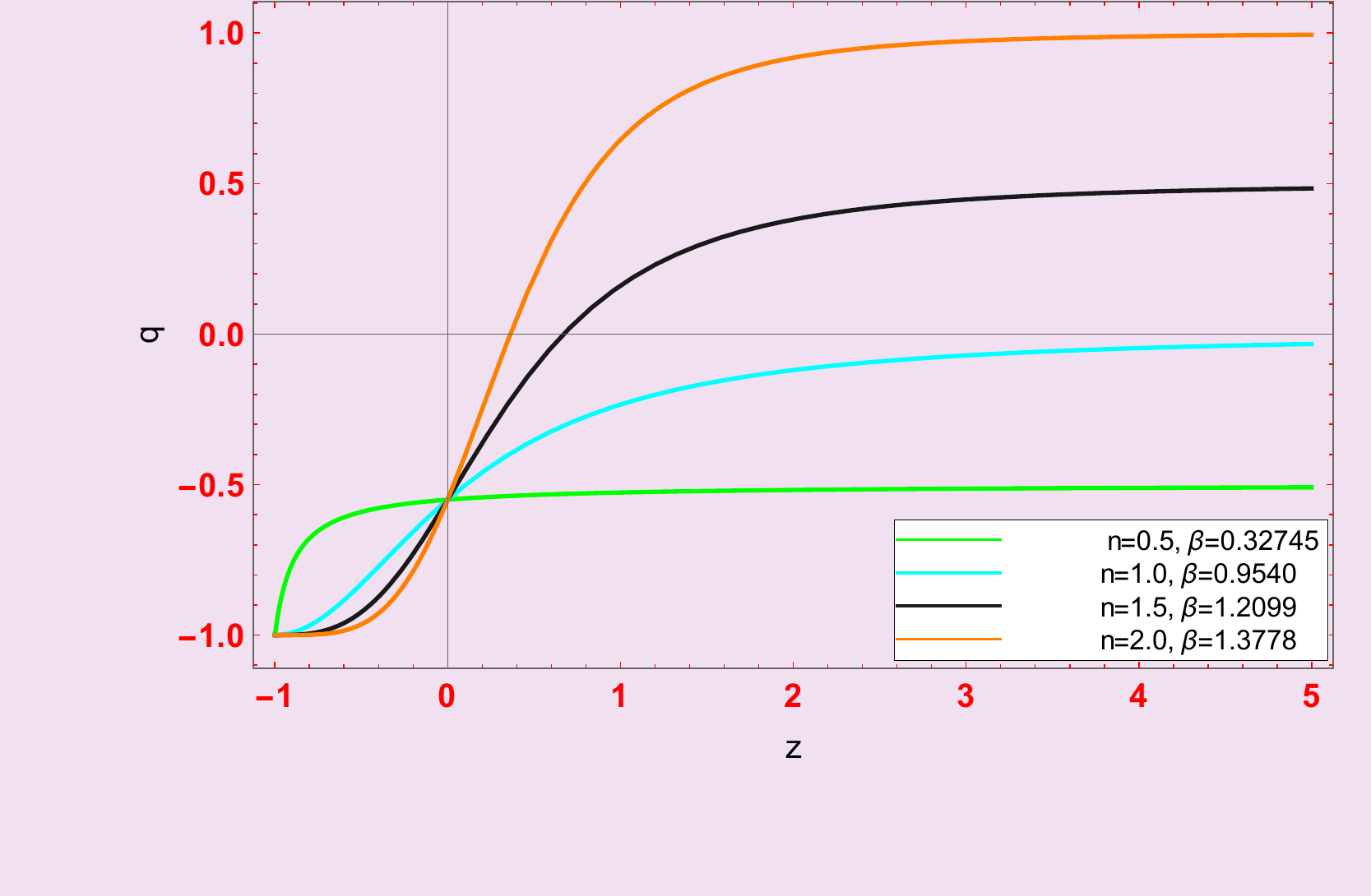} \\ 
\mbox (a) & \mbox (b)%
\end{array}%
$%
\end{center}
\caption{\scriptsize The plots of Hubble parameter $\frac{H}{H_0}$ and deceleration parameter $q$ for the model.}
\end{figure}

In the current study, we are keen on exploring the multiple regimes, or the phase change from a decelerated to an accelerated expansion by restricting a model parameter $ n $. Different choices of $n$ provide different values of the model parameter $\beta$ which is shown in the above Table 1.\\

For the considered choice of scale factor $a(t)$ in Eq. (\ref{15}), the model parameter $n$ have already been constrained in our previous work (\cite{ritzannals}) using some observational datasets: $SNeIa$ dataset, $H(z)$ (Hubble parameter) dataset, $BAO$ (Baryon Acoustic Oscillation data) and their combinations as joint observational datasets $H(z) + SNeIa$ and $H(z) +SNeIa+BAO$ and found out the best-fit values of the model parameters mentioned in Table 2. As we can observe that the constrained value of $n$ is approximately equal to $1.5$, so in this work among the three chosen value of $n$, the value corresponds to $n=1.5$ will provide the best outcomes for the further analysis in sections 5, 6 and 7.\\

\begin{table}
\caption{Outcomes of data analysis for our model} 
\begin{center}
\label{table2}
\begin{tabular}{l c c c r} 
\hline\hline
\\ 
{Data} \,\,\,\,\,  &  \,\,\,\, $ \chi^2_{min} $ \,\,\,\,\,  &  \,\, \, $ H_0 $  \,\,\,\,\,   &  \,\,\,\,\, $  n $ \,\,\,\\ 
\\
\hline 
\\
{$ H(z) $}   \,\,\,\,\,  &  \,\, \,   $ 19.5872 $   \,\,\,\,\,  &  \,\, \,   $ 68.0000 $   \,\,\,\,\,  &  \,\, \,   $ 1.5176 $
\\
\\
{$ SNeIa $}   \,\,\,\,\,  &  \,\, \,    $ 565.5074 $   \,\,\,\,\,  &  \,\, \,   $ 66.3402 $   \,\,\,\,\,  &  \,\, \,   $ 1.5907 $ 
\\
\\
{$ BAO $}   \,\,\,\,\,  &  \,\, \,   $ 2.1622 $   \,\,\,\,\,  &  \,\, \,   $ 66.4138 $   \,\,\,\,\,  &  \,\, \,   $ 1.5009 $ 
\\
\\
{$ H(z) $ + $ SNeIa $ }  \,\,\,\,\,  &  \,\, \,  $ 589.7361 $  \,\,\,\,\,  &  \,\, \,  $ 65.6148 $  \,\,\,\,\,  &  \,\, \,  $ 1.5396 $
\\
\\
{$ H(z)$ + $ SNeIa $ + $ BAO $}   \,\,\,\,\,  &  \,\, \,   $ 597.2995 $   \,\,\,\,\,  &  \,\, \,   $ 63.0579 $   \,\,\,\,\,  &  \,\, \,   $ 1.5060 $ 
\\
\\ 
\hline\hline  
\end{tabular}    
\end{center}
\end{table}

Inflation in the Universe depends on the sign of $ q $, according to Eq. (\ref {18}). The plot of $H$ and $q$ \textit{w.r.t} $z$ in FIG. 1(a), 1(b) in our model exhibit expanding Universe,  acceleration (\textit{i.e.} $ q<0 $) and phase transition (\textit{i.e.} $ q=0 $) from deceleration to acceleration according as the choices of $n$ and $\beta$ respectively. It is widely accepted that the Universe has had an accelerating phase in recent times, hence it must have had a steady expansion in the past. It is seen from the plot of $q$, FIG. 1(b) that for $n=0.5,1$, the model predicts no phase transition in the Universe or rather it is ever accelerating, which is not compatible with the standard model of cosmology as it is the belief that the early deceleration is pledged for the structure formation in the Universe. Also, exceeding the value of $n$ (\textit{i.e.} $n=1.5, 2$) allows the model of the Universe to anticipate the transition from deceleration to acceleration at $z_{transition}=0.6710,0.3627$ respectively. As the value of $n$ increases, phase transition $z_{transition}$ shifts towards zero. In our model, phase transition ($q = 0$) will occur at the time $z = -1 + (\frac{sinh({\beta})^2} {n-1})^{\frac{1}{2n}}$, here any value of $n <= 1$ will not give feasible value of $\beta$ for possible phase transition, so phase transition is possible when $n > 1$. Henceforth, we will consider the choice of $n>1$ which predicts the phase transition in the Universe and is well consistent with standard cosmology. Next, we emphasize on the physical parameters of the model for the flat Universe only as the current observations are in the favor of $k=0$ so that $\Omega_{k} \rightarrow 0$.\newline

\cite{ritzepjc2}, \cite{ritzepjc1}, \cite{ritzannals}, \cite{ritzepjp} and \cite{ritzmpla} addressed the specific dynamical behaviour of these geomertical parameters. In this study, we attempt to explore the physical dynamics of a DE interacting model using this parametrization. We obtain the physical parameters as follows using equations (\ref{17}) and (\ref{18}).
\begin{equation}
\Omega _{R}=\Omega _{0,R}\frac{(1+z)^{4}}{(1+z)^{2n}(sech(\beta))^2+(tanh(\beta))^2}\text{,}
\label{20}
\end{equation}%
\begin{equation}
\Omega _{B}=\Omega _{0,B}\frac{(1+z)^{3}}{(1+z)^{2n}(sech(\beta))^2+(tanh(\beta))^2 }\text{,}
\label{21}
\end{equation}%
\begin{equation}
\Omega _{DM}=\Omega _{0,DM}\frac{(1+z)^{3(1-\epsilon)}}{(1+z)^{2n}(sech(\beta))^2+(tanh(\beta))^2}%
\text{,}  \label{22}
\end{equation}

\begin{equation}
\Omega _{DE}=\left( 1+\Omega _{k}\right) -\frac{\left[ \Omega _{0,R}\left(
1+z\right) ^{4}+\Omega _{0,B}\left( 1+z\right) ^{3}+\Omega _{0,M}\left(
1+z\right) ^{3-3\epsilon }\right] }{(1+z)^{2n}(sech(\beta))^2+(tanh(\beta))^2}\text{.}  \label{23}
\end{equation}%

Equation (\ref{23}) is verified with $\Omega _{0,DE}=1+\Omega _{k}-(\Omega_{0,R}+\Omega _{0,B}+\Omega _{0,DM})$. 
Equations (\ref{14c}), (\ref{14d}) and (\ref{14e}) can be written with the help of equations (\ref{17}) and (\ref{18}) as, 
\begin{equation}
\frac{p_{DE}}{H_{0}^{2}}=\frac{1}{8\pi G}\big(\big[2n\big(1-tanh^2\big(sinh^{-1}\big(\frac{sinh(\beta t_0)}{(1+z)^n}\big)\big)\big)-3\big]-\Omega_{k}\big)\big((1+z)^{2n} sech(\beta)^2 \big(1+\frac{(sinh(\beta))^2}{(1+z)^{2n}} \big)-\Omega
_{0,R}\left( 1+z\right) ^{4}\text{,}  \label{24}
\end{equation}

\begin{equation}
\frac{\rho _{DE}}{H_{0}^{2}}=\frac{3}{8\pi G} \big[(1+z)^{2n}(sech(\beta))^2 \big( 1+ \big(\frac{sinh(\beta)^2}{(1+z)^{2n}} \big) \big) (1+\Omega_{k})-\big(\Omega_{0,R} (1+z)^4 +\Omega_{0,B}(1+z)^3+\Omega_{0,{DM}}(1+z)^{3-3\epsilon} \big) \big] \text{,}  \label{25}
\end{equation}%
and

\begin{equation}
\omega _{DE}=\frac{1}{3}\frac{\big(\big[2n\big(1-tanh^2\big(sinh^{-1}\big(\frac{sinh(\beta t_0)}{(1+z)^n}\big)\big)\big)-3\big]-\Omega_{k}\big)\big((1+z)^{2n} sech(\beta)^2 \big(1+\frac{(sinh(\beta))^2}{(1+z)^{2n}} \big)\big)-\Omega
_{0,R}\left( 1+z\right) ^{4}}{\big[(1+z)^{2n}(sech(\beta))^2 \big( 1+ \big(\frac{sinh(\beta)^2}{(1+z)^{2n}} \big) \big) (1+\Omega_{k})-\big(\Omega_{0,R} (1+z)^4 +\Omega_{0,B}(1+z)^3+\Omega_{0,{DM}}(1+z)^{3-3\epsilon} \big) \big] }  \label{26}
\end{equation}

\begin{figure}[H]
\begin{center}
$%
\begin{array}{c@{\hspace{.1in}}cc}
\includegraphics[width=3.5 in, height=2.5 in]{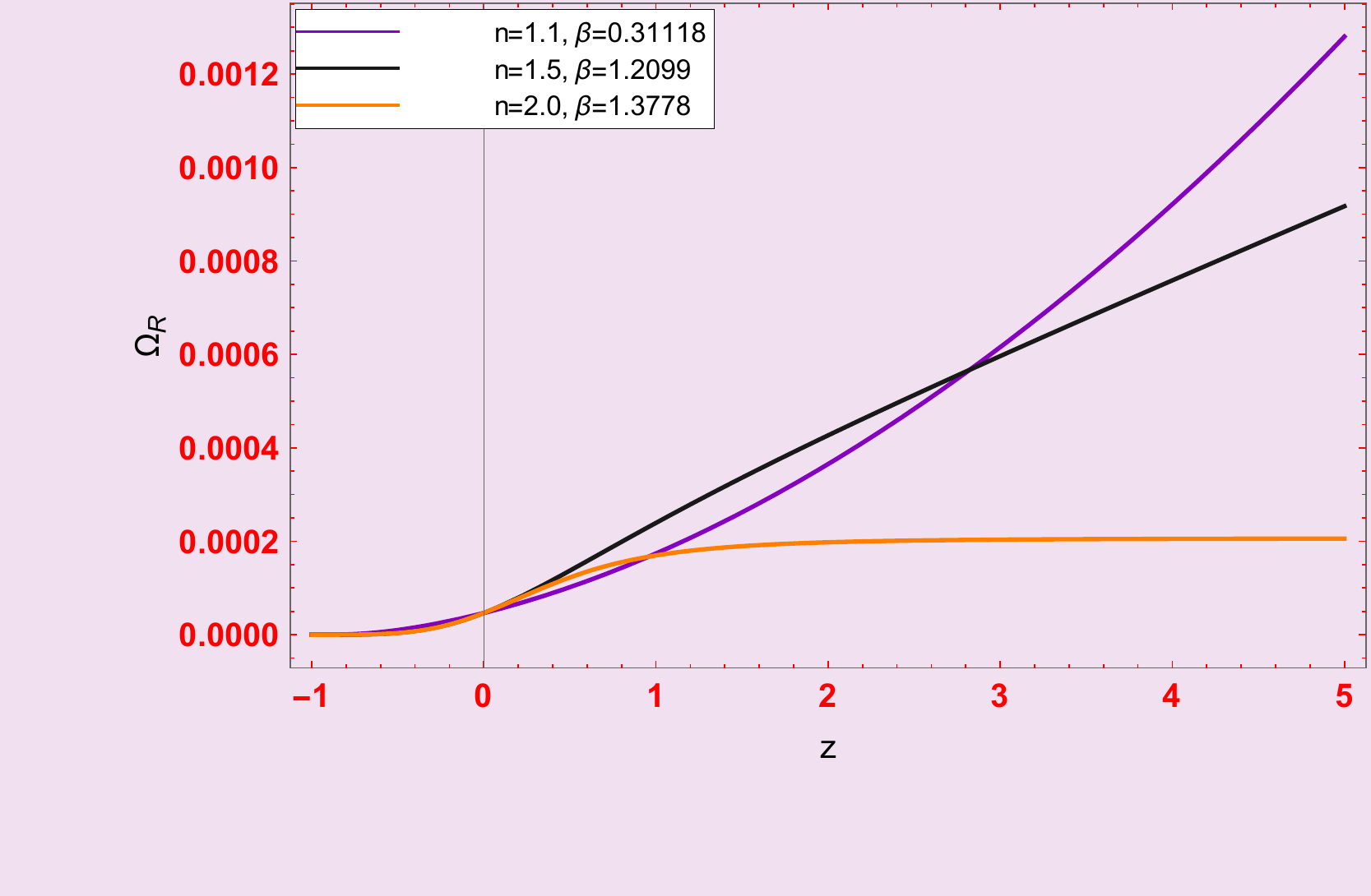} & %
\includegraphics[width=3.5 in, height=2.5 in]{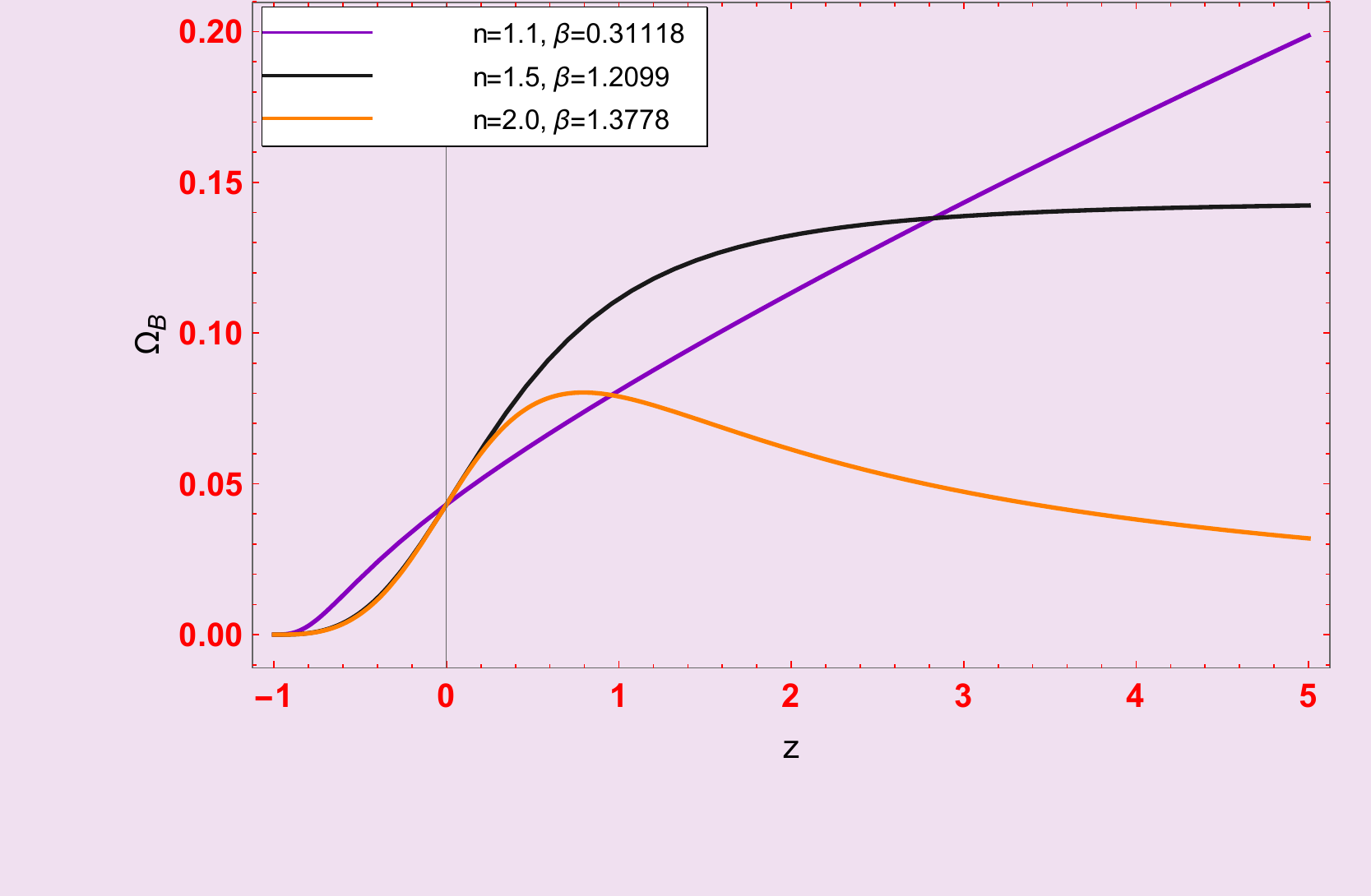} \\ 
\mbox (a) & \mbox (b)%
\end{array}%
$%
\end{center}
\caption{\scriptsize The plots of density parameters of radiations $\Omega_{R}$ and baryons $\Omega_{B}$ for the model.}
\end{figure}

\begin{figure}[H]
\begin{center}
$%
\begin{array}{c@{\hspace{.1in}}cc}
\includegraphics[width=3.5 in, height=2.5 in]{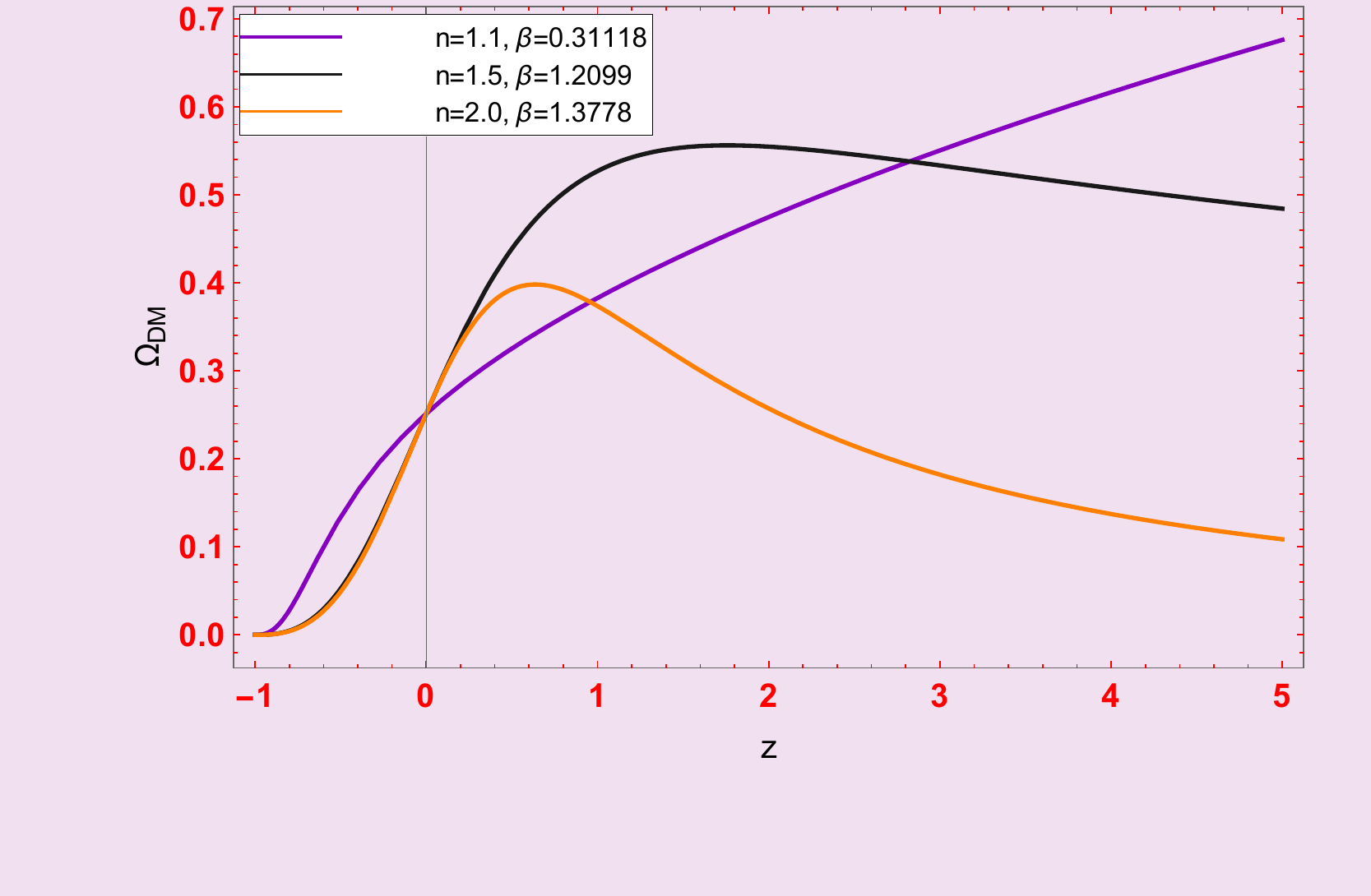} & %
\includegraphics[width=3.5 in, height=2.5 in]{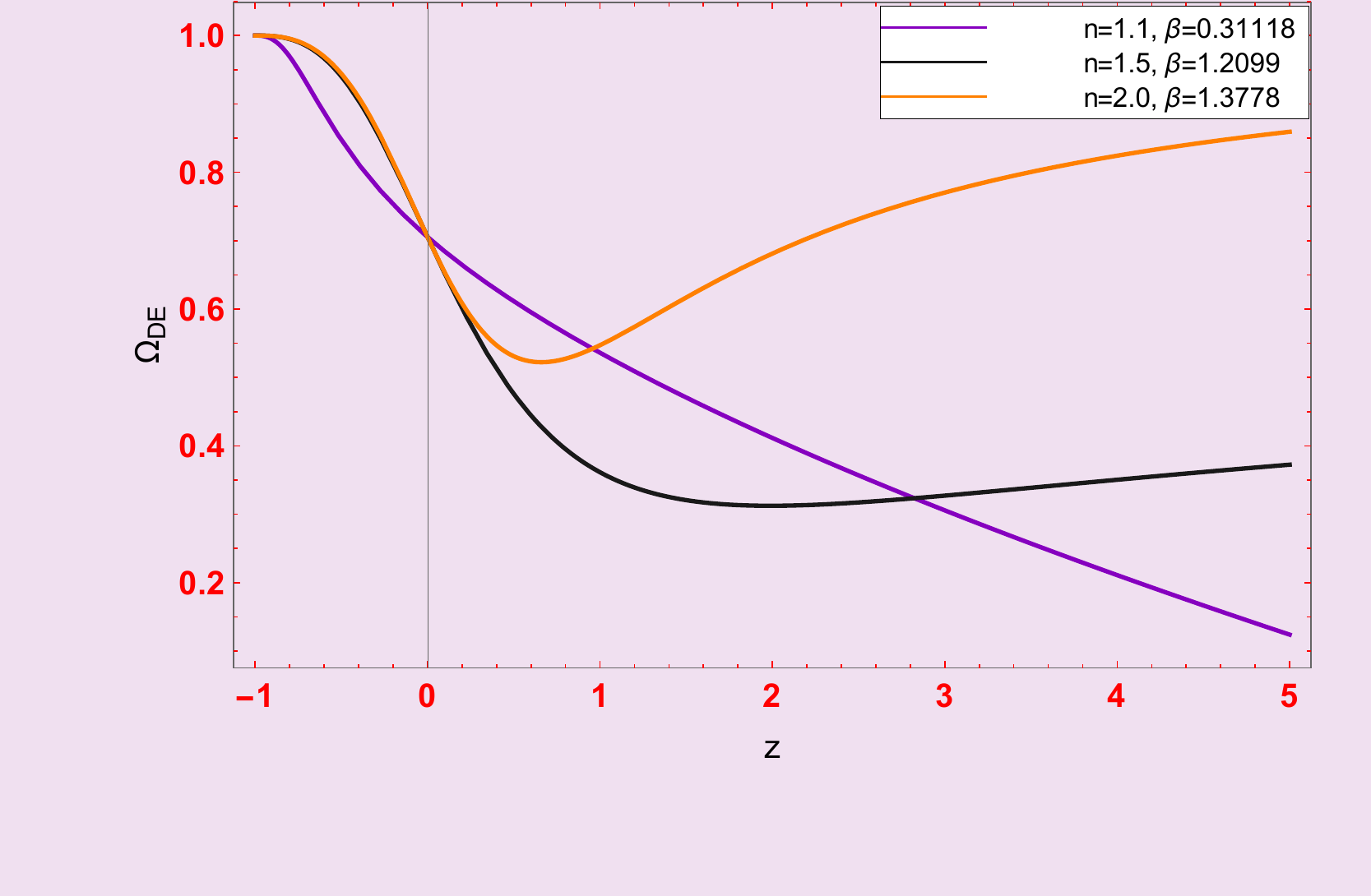} \\ 
\mbox (a) & \mbox (b)%
\end{array}%
$%
\end{center}
\caption{\scriptsize The plots of density parameters of dark matter $\Omega_{DM}$ and dark energy $\Omega_{DE}$ for the model.}
\end{figure}

The contrasted behavior of the different components of the density parameter ($\Omega_{R}$, $\Omega_{B}$, $\Omega_{DM}$ and $\Omega_{DE}$) with respect to redshift $z$ are shown in the above figures FIG. 2(a), 2(b) and FIG. 3(a), 3(b) . It is clear that during the beginning of the universe's evolution, the density parameter of radiation and matter (baryonic $+$ dark matter) dominates over the density parameter of DE. In the late times, $\Omega_{DE}$ have been monotonically increasing whereas the density parameters of $\Omega_{R}$, $\Omega_{B}$ and $\Omega_{DM}$ have been monotonically falling for different values of $n$. Additionally, at $z=-1$, all the density parameters $\Omega_{R}$, $\Omega_{B}$ and $\Omega_{DM}$ abruptly collapsed to $zero$, whereas $\Omega_{DE}$ access to $1$. Additionally, the sum of all values of density parameters adds up to unity which is consistent with many observational findings.

\begin{figure}[H]
\begin{center}
$%
\begin{array}{c@{\hspace{.1in}}cc}
\includegraphics[width=3.5 in, height=2.5 in]{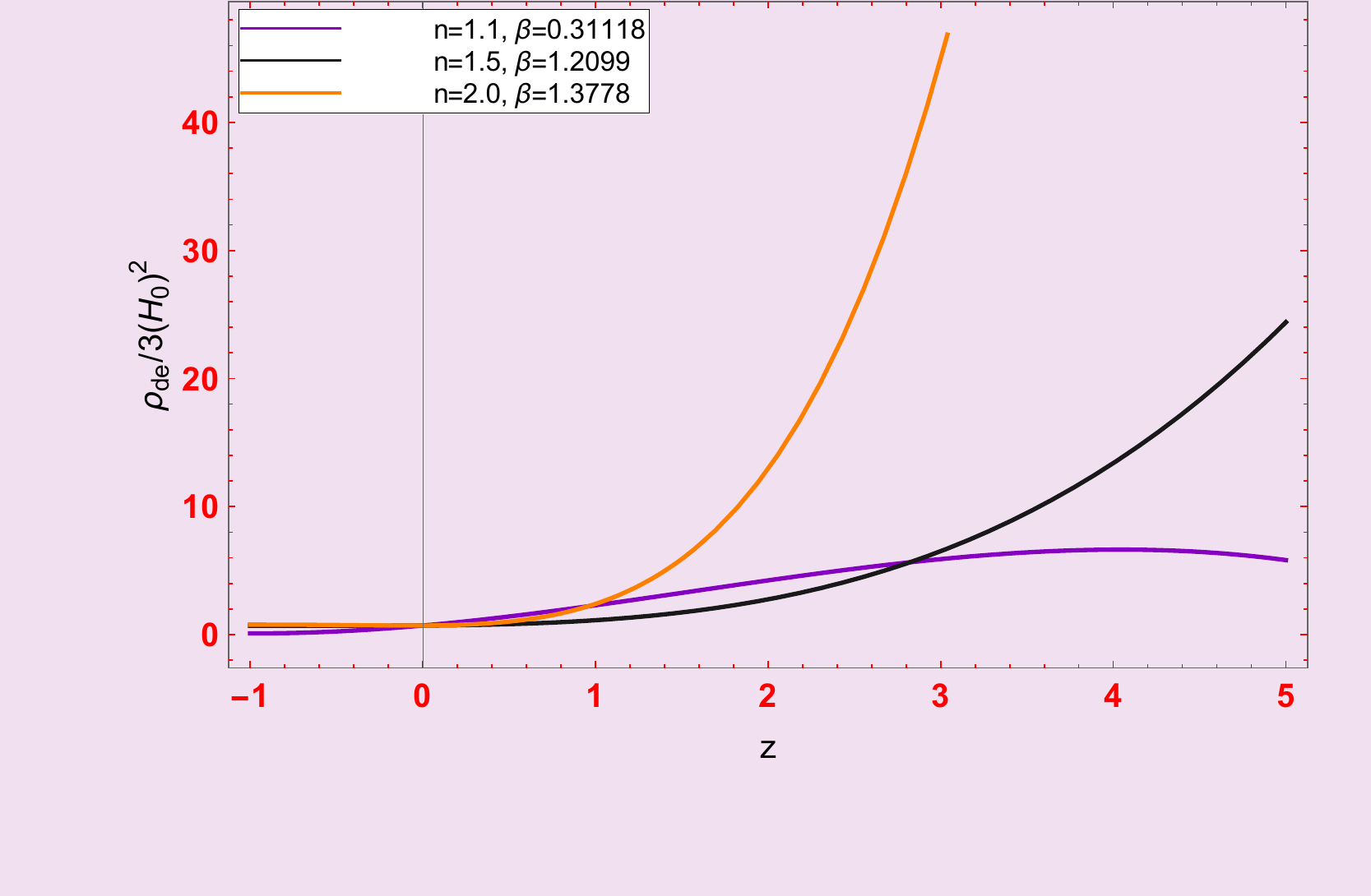} & %
\includegraphics[width=3.5 in, height=2.5 in]{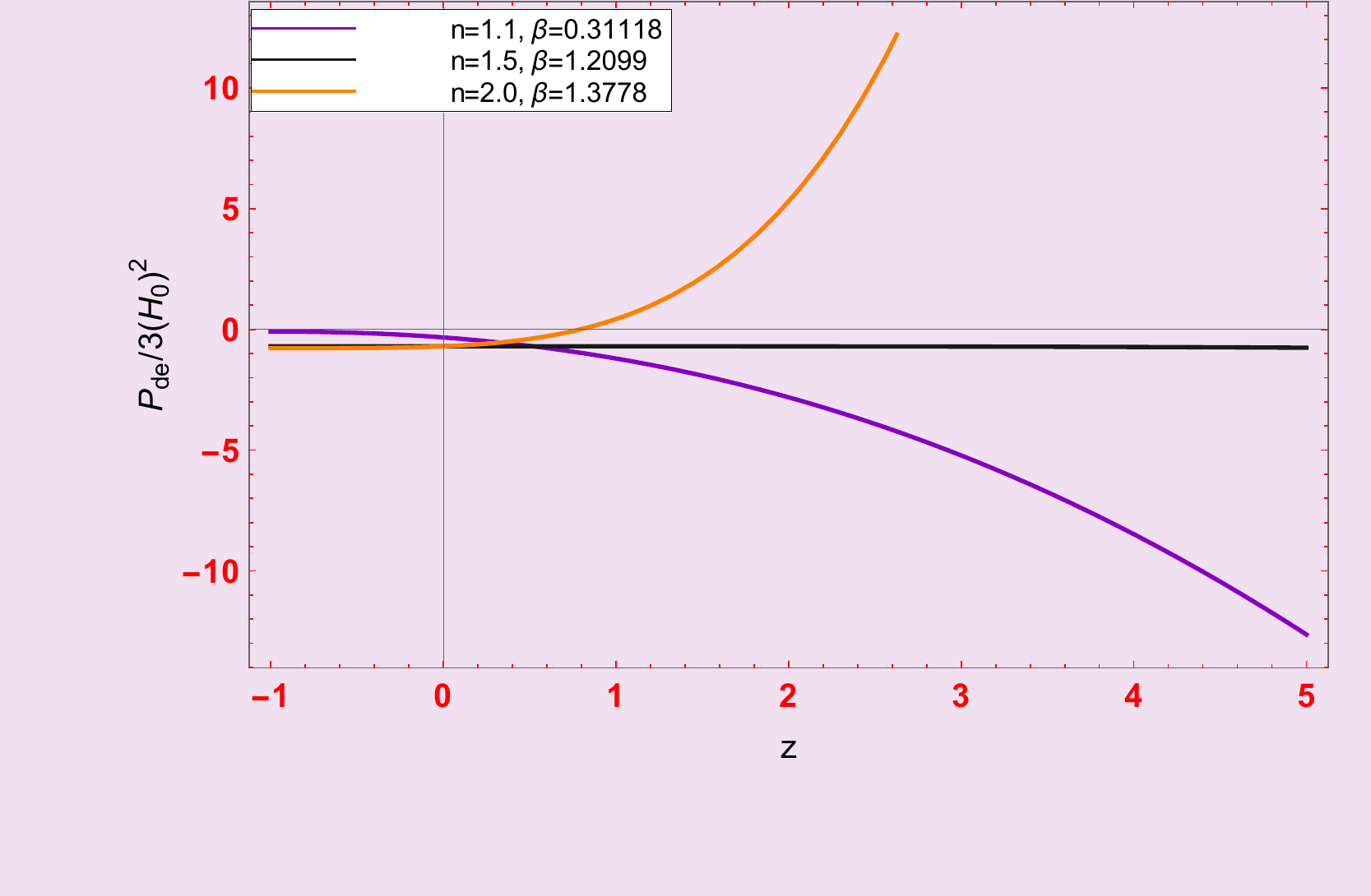} \\ 
\mbox (a) & \mbox (b)%
\end{array}%
$%
\end{center}
\caption{\scriptsize The plots of energy density of dark energy $\rho_{de}$ and pressure of dark energy $p_{de}$ for the model.}
\end{figure}

We display the curves of DE's energy density $\rho_{DE}$ and pressure of DE $P_{DE}$ with respect to redshift $z$ for all selected model parameter values $n$ and $\beta$ taken from Table 1  in the FIG. 4(a) and 4(b). For the changing model parameter $n$, the slope of the curve $\rho_{DE}$ decreases. As a result, we observe that the energy density of the scalar field is initially high during the Universe's evolution and finally approaches $0$, indicating that the amount of $\rho_{DE}$ decreases to almost $0$ in later eras.
\begin{figure}[H]
\begin{center}
$%
\begin{array}{c@{\hspace{.1in}}cc}
\includegraphics[width=3.5 in, height=2.5 in]{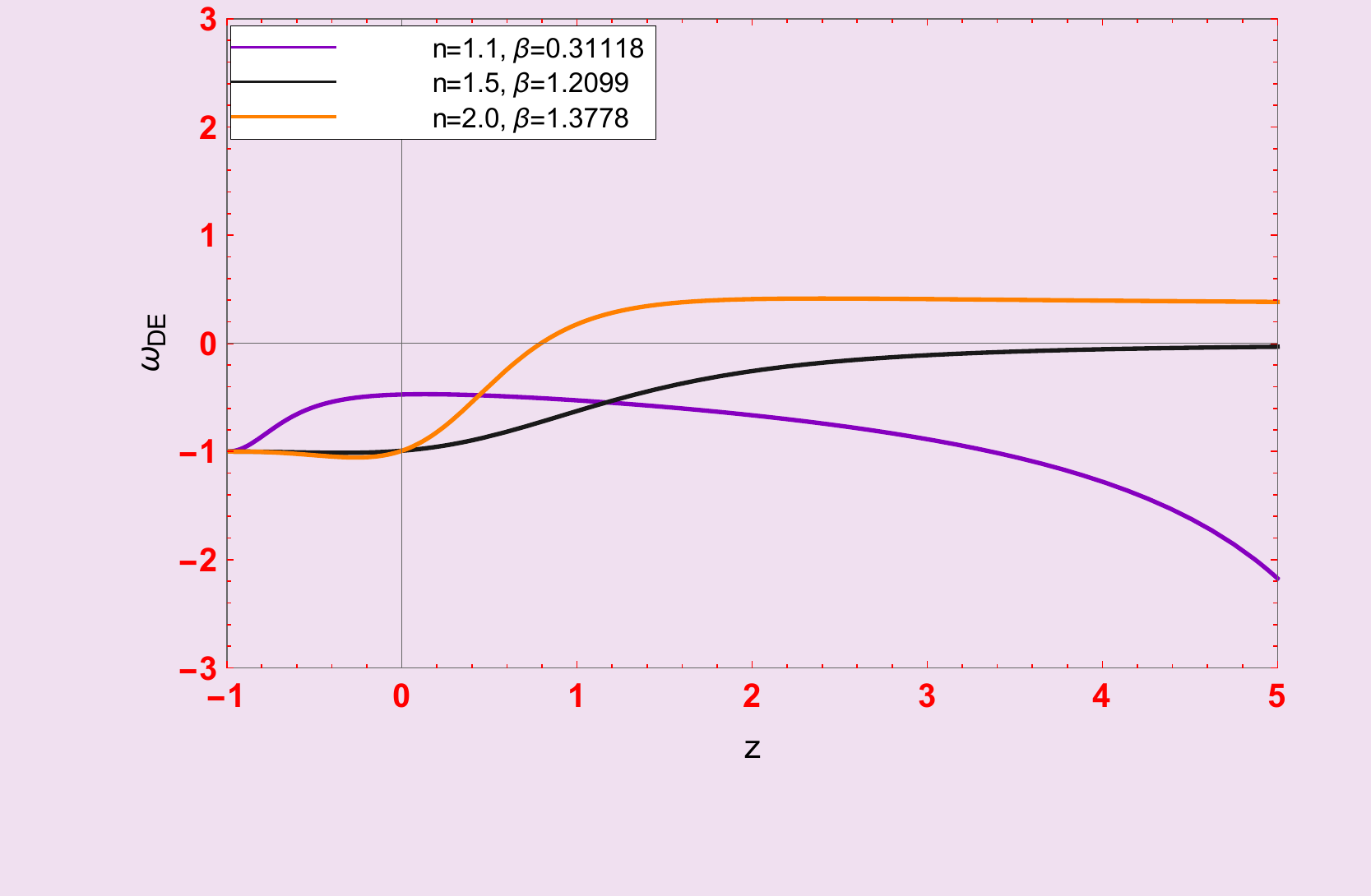} 
\end{array}%
$%
\end{center}
\caption{\scriptsize The plots of EoS parameter $\omega_{de}$ for the model.}
\end{figure}

As it is also self-evident, the behavior of the EoS parameter $\omega$ can be examined to understand many cosmic evolution stages of the Universe. We have talked about different energy densities and pressures in our model represented in Eqs (\ref{2}), (\ref{3}). The three trajectories of the EoS parameter of DE $\omega_{DE}$ for various options of $n$ listed in Table 1 are shown in FIG. 5. The current value of $\omega_{DE}$ for $n=1.1$ is $\omega_{0,{DE}}= -0.472614$ whereas $\omega_{0,{DE}} \approx -0.992394, -0.992808 $ \textit{i.e.} $(\omega_{0,{DE}}\approx -1)$ as we increase the value of $n$, the EoS parameter becomes consistent with the constraint range provided by Riess \cite{agriess}.

\section{Energy conditions}
In general relativity, there are several common energy conditions (ECs) that impose limitations on the ability of some regions to have negative energy density.
In other words, ECs are accepted as a valid generalization of the energy-momentum tensor \cite{carr}, \cite{san1}, \cite{san2}, \cite{sen}, \cite{bert}, \cite{gar} to the entire Universe, where energy density can never be negative. Using ECs, a wide range of significant singularity problems involving wormholes, black holes, and other phenomena are thoroughly studied. The ECs can primarily be described in two ways: (i) geometrically, where ECs are well characterized in terms of Ricci tensor or Weyl tensor, and (ii) physically, where ECs are expressed as a function of the physical world. Following is a definition of the ECs.

\begin{itemize}
\item Null energy condition (NEC) $\Leftrightarrow$ $ \rho_{Total}+p_{Total} \geq 0 $,
\item Weak energy condition (WEC) $\Leftrightarrow$ $ \rho_{Total} \geq 0 $, $ \rho_{Total}+p_{Total} \geq 0 $,
\item Strong energy condition (SEC) $\Leftrightarrow$ $ \rho_{Total}+3p_{Total} \geq0 $,
\item Dominant energy condition (DEC) $\Leftrightarrow$ $ \rho_{Total} \geq 0 $ , $ \rho_{Total} \geq |p_{Total}| $ .
\end{itemize}
Furthermore, the potential energy ($V(\phi)$) and scalar field ($\phi$) of the model can be used to express the energy density ($\rho$) and isotropic pressure ($p$), respectively. In GR, the point-wise ECs are thus defined as:
\begin{itemize}
\item NEC $\forall \, V(\phi)$,
\item WEC $\Leftrightarrow$ $V(\phi) \geq \frac{\dot{\phi}^2}{2}$,
\item SEC $\Leftrightarrow$ $V(\phi) \leq \dot{\phi}^2$,
\item DEC $\Leftrightarrow$ $V(\phi) \geq 0.$
\end{itemize}

Cosmology relies heavily on ECs to comprehend the behaviour of the cosmos and all of its varied characteristics. The type of interaction between different elements in the interacting theory depends on ECs. For example, WEC prevents negative energy densities in matter. The stability of matter must be comprehended because of this condition. If the ECs in the interaction theory are violated, new mechanics outside of the standard model may arise. This can be done by using the ECs of an interacting theory to control how different components interact with one another. It is possible to impose restrictions on the interactions between different parts of an interacting theory using its ECs. The most common representation of DE is a scalar field with a potential energy density that is either constant or slowly varies over time. Dark matter, on the other hand, is described as a non-relativistic fluid that reacts to gravity but not to itself or to other particles. One way to study this interaction is to examine the behaviour of the scalar field $\phi$ that describes DE and how it interacts with the DM.  The energy conditions are essential to this study because it sets limitations on the scalar field's behavior and possible energy density. In conclusion, the scalar field $\phi$ and ECs are crucial tools for understanding how DE and DM interact in terms of their energy densities.\\

Another key subject discussed here is the analysis of various ECs to ascertain whether weak, null, strong, and dominating ECs are satisfied in the scenarios offered. In order to investigate different ECs for our model, we plot some data for interacting scenarios.

\begin{figure}[H]
\begin{center}
$%
\begin{array}{c@{\hspace{.1in}}cc}
\includegraphics[width=3.2 in, height=3.2 in]{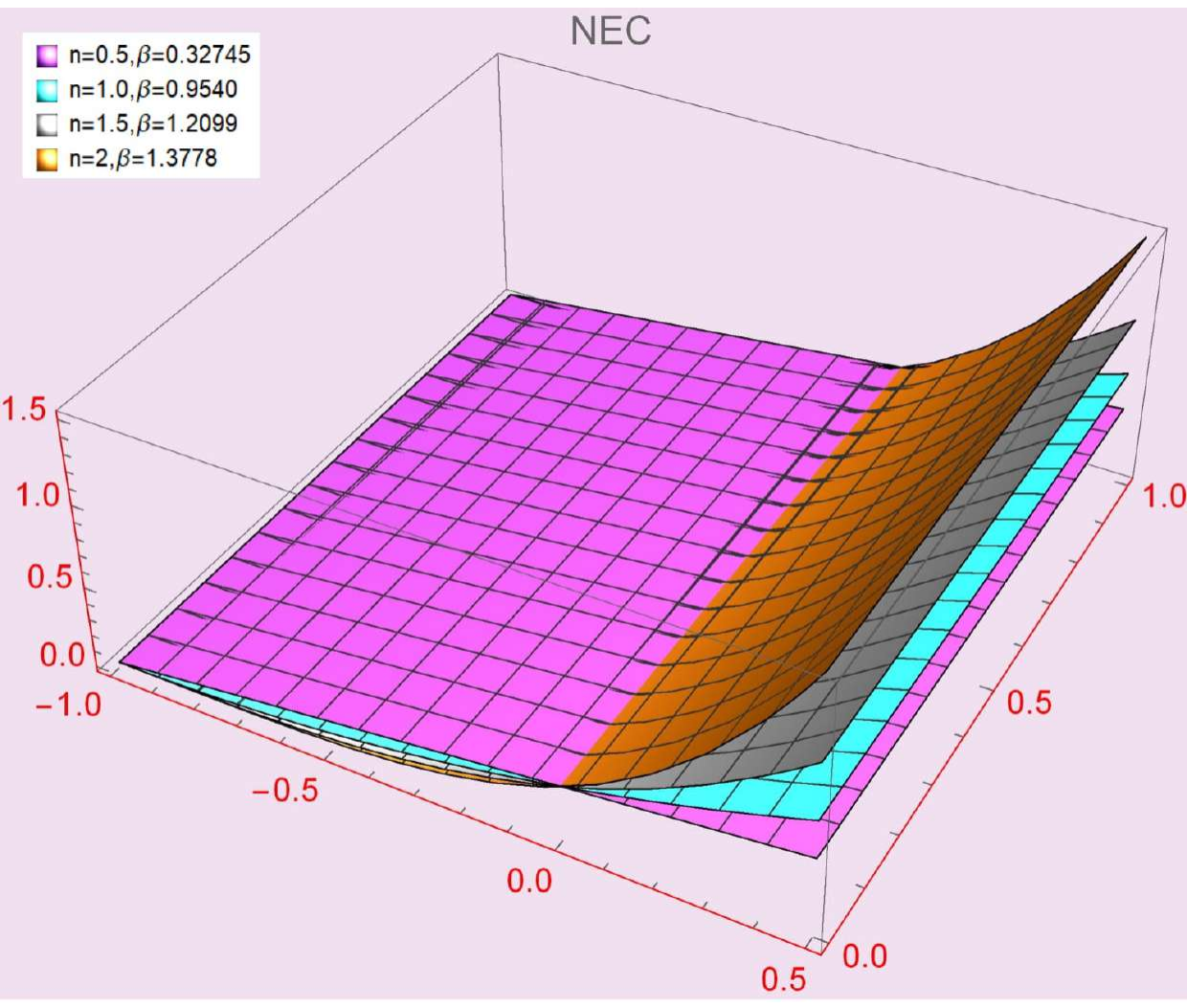} & %
\includegraphics[width=3.2 in, height=3.2 in]{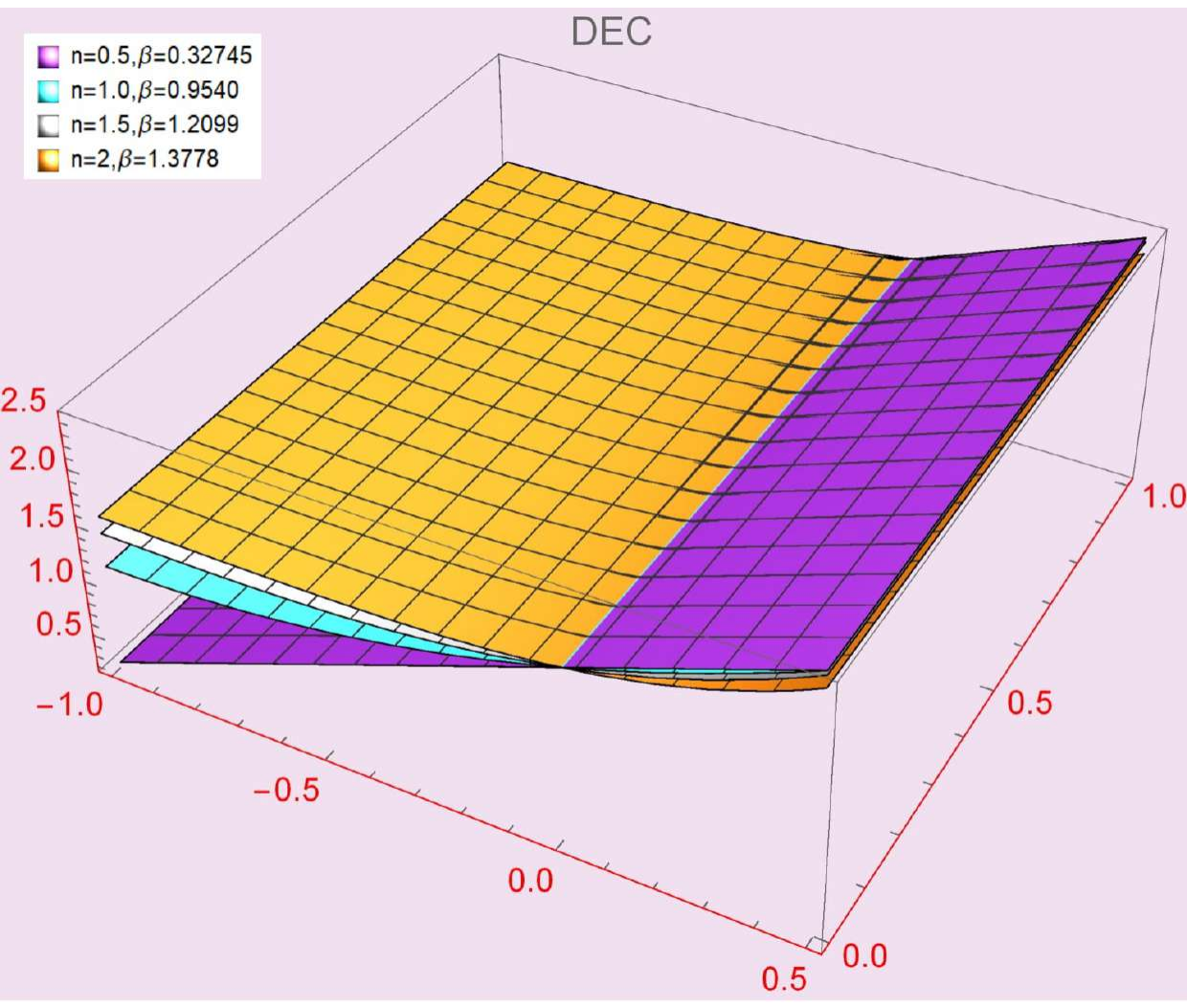} \\ 
\mbox (a) & \mbox (b) %
\end{array}%
$%
\end{center}
\caption{\scriptsize The plots of $NEC$ and $DEC$ for the model.}
\end{figure}

\begin{figure}[H]
\begin{center}
$%
\begin{array}{c@{\hspace{.1in}}cc}
\includegraphics[width=3.5 in, height=3.2 in]{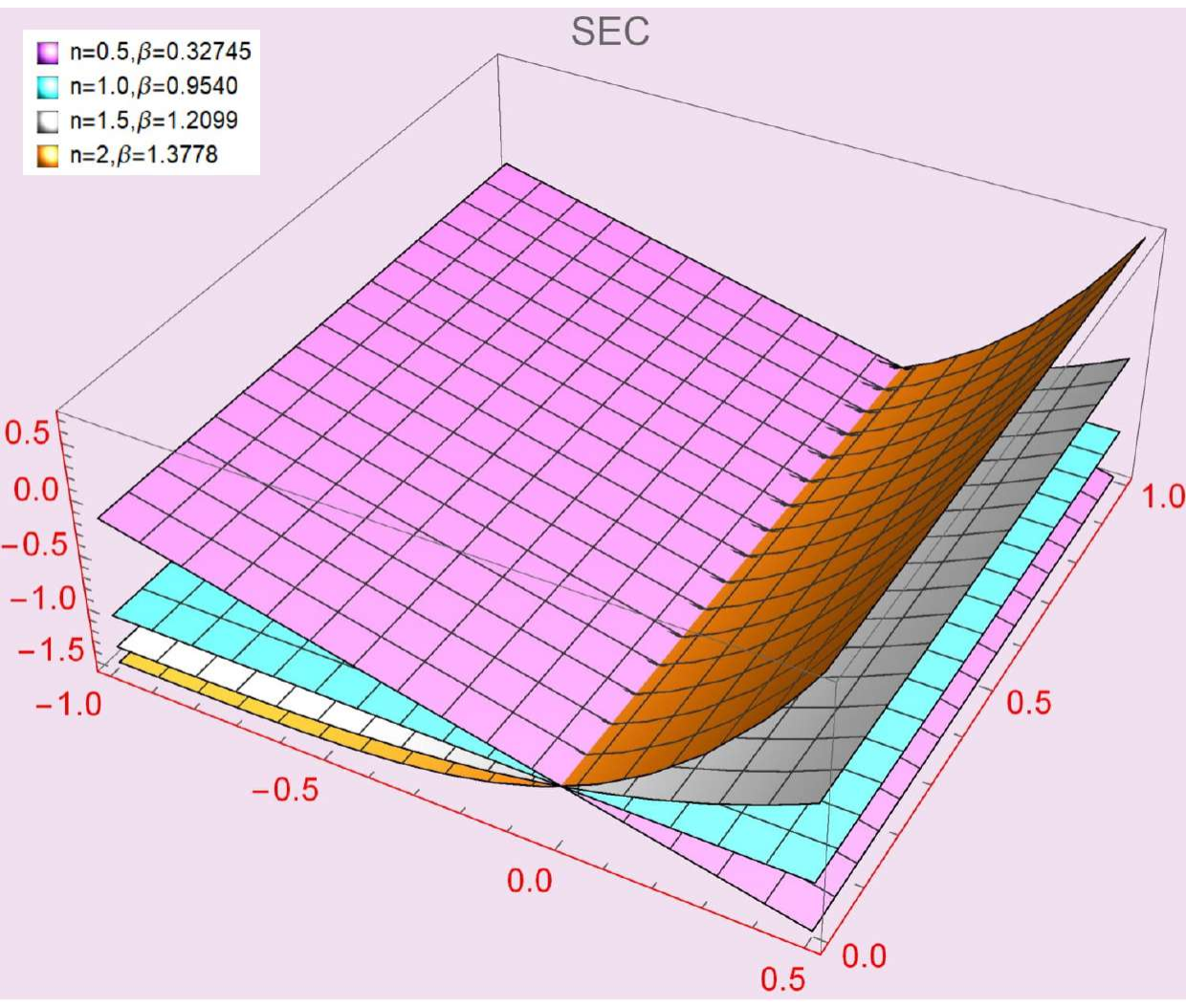} \\ 
\end{array}%
$%
\end{center}
\caption{\scriptsize The plot of $SEC$ for the model.}
\end{figure}

We can see from the above FIG. 6(a), 6(b) that while $NEC$ and $DEC$ are satisfied in our model whereas $SEC$ in FIG. 7 is not, which results in our model showing the Universe's rapid expansion. However, the $SEC$ violation is similar to the Universe having some exotic matter to cause anti-gravitational effects. As a result, the Universe's expansion is accelerating under the influence of an apparent strong enough repulsive force.

\section{Velocity of sound}
As is well known, one of the most important tests for a cosmological model's viability is its stability under linear perturbations. However, a stringent restriction is imposed by the requirement that the speed of sound ($C_s^2$) must be sufficiently less than $1$ to prevent unwanted oscillations in the matter power spectrum. Here, we use a choice of the appropriate parameters to plot the sound velocity ($C_s^2$) graph for our derived models, as shown in the following figure:
\begin{figure}[H]
\begin{center}
$%
\begin{array}{c@{\hspace{.1in}}cc}
\includegraphics[width=3.5 in, height=2.5 in]{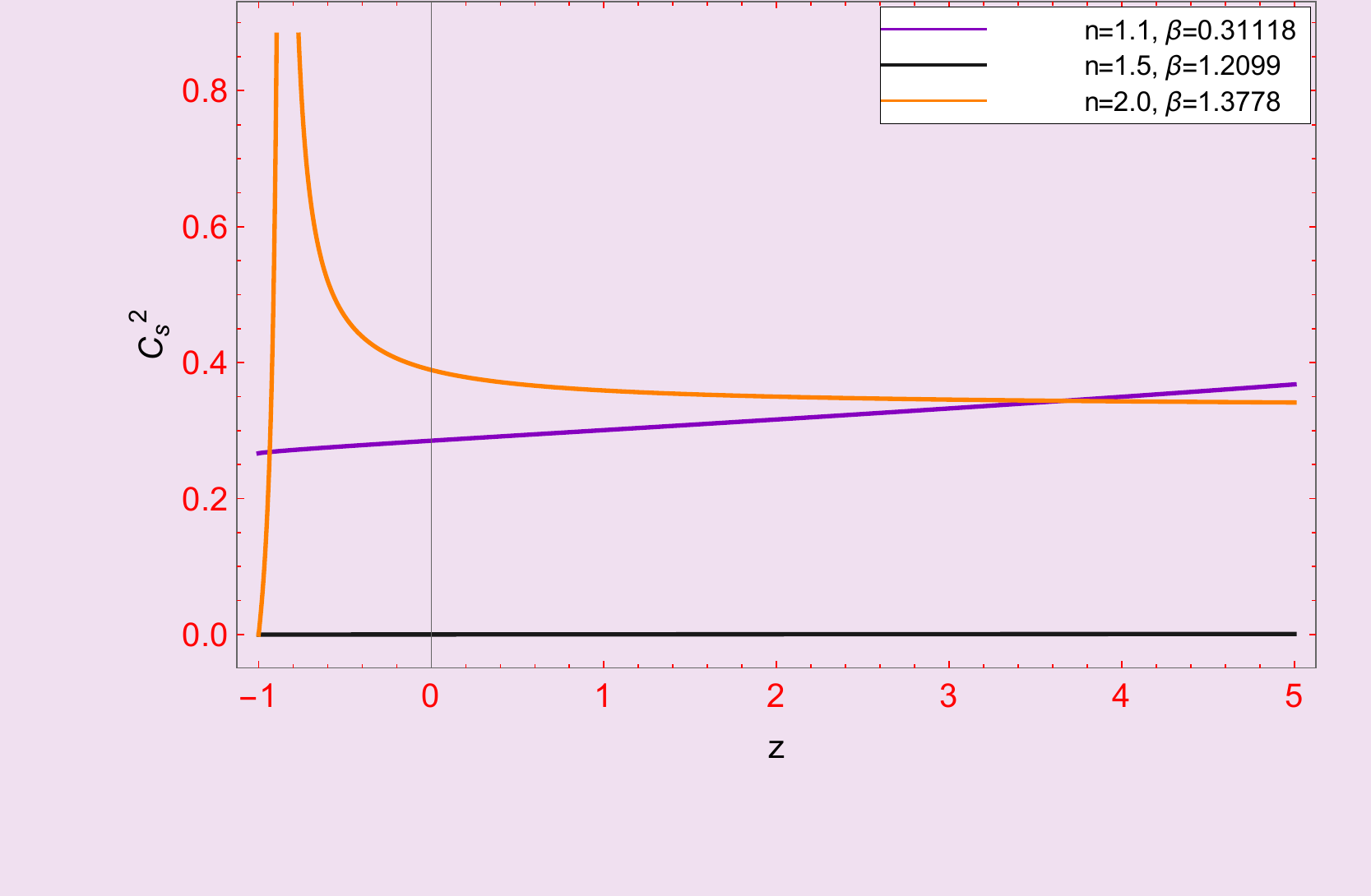} \\ 
\end{array}%
$%
\end{center}
\caption{\scriptsize The plot of $C_s^2$ for the model.}
\end{figure}
The plot of velocity of sound $C_s^2$ in FIG. 8 exhibits the stability in our interacting model for considered choices of $n$.\\

The sound speed in the given DM-DE scenario is an important tool that can help us to understand the dynamics of Universe by studying the perturbation in the energy density and pressure of matter and energy, which would signal on the stability or instability in the system. In our model, the sound speed is independent of the coupling parameter $\epsilon$. The model is always stable for any value of coupling interaction parameter and independent of its sign and magnitude.

\section{Swampland conjecture}
Quantum gravity theory, which can clearly distinguish between the effective field models and the nominal string theory, is becoming more and more popular in the modern era as it is being used to link modern cosmological models. The Swampland criteria have just been established to select field potentials appropriately. We will therefore clarify the connection between DE and the refined swampland program conjecture in this section. As is well known, the swampland conjecture sheds light on many cosmological theories as well as the physics governing black holes, inflation, and other phenomena. Interest in the relationship between real-world cosmological models and gravity's quantum theory, which can distinguish between the effective field models connected to the nominal string landscape, has recently increased. Recently, swampland criteria were developed to aid in choosing field potential candidates. Excitement was notably generated by the fact that accurate de Sitter solutions with non-negative cosmological constants do not fit with the string landscape, which makes it challenging to connect high-energy structures to fundamental theories.\newline

Recent work \cite{ooguri} has determined the swampland criterion of the string theory bound using the DE quintessence model. To connect DE quintessence models, many works, including \cite{benetti}, \cite{david}, \cite{hertz}, \cite{brahma}, \cite{raveri}, and \cite{agrawal} approached the Swampland criterion after the work in \cite{ooguri}. The researchers recently investigated the Swampland conjecture regarding the validity of the alternative theory of gravity (f(R) theory, f(R,T)) \cite{pradhan9}. The Swampland constraints impose restrictions and control on a field theory that defies quantum gravity \cite{odinstov1}. Although these theories are still in their infancy and solve a few problems, they are altering and evolving quickly in particular circumstances, such as warm inflation. In light of this, we decide to link these concepts to DE. In later works, we shall try to expand on this idea and establish a more direct link between the swampland program, DM, and DE and in action. The nature of the idea is still unknown, even though researchers have taken into account various DE models and examined their cosmological applications. \\

In this article, we look into the cosmological limitations imposed by two-string Swampland criteria. For $V > 0$, these requirements include a lower constraint on $\frac{\phi V} {V}$ and an upper limit on the range that scalar fields can cover. These two requirements appear to have the fundamental drawback of generally contradicting inflationary theories. Applying the same standard to DE in the present period, we find that some quintessence models can satisfy these constraints simultaneously. If the two Swampland conditions are fulfilled, then the cosmos will undergo a phase change within a few Hubble times, according to our theory. In light of this justification, we propose the improved swampland conjecture in the format below. \\
Conjecture 1. $\frac{|V'|}{V}>C_{1} M_{pl}^2$ \,\,\, and \,\,\, Conjecture 2. $\frac{|V''|}{V}<-C_{2}M_{pl}^2$.

As we have already observed, the idea is to fill them with an unusual sort of matter that meets the inequality $1+3\omega<0$ to predict the acceleration in the Universe. The measurements show that the energy responsible for the acceleration satisfies the condition $\omega \simeq -1$. Since the formation of the EoS parameter $\omega$ for our model was already discussed in earlier sections, we will now take into consideration the following equations to obtain a matter field that exhibits peculiar behavior and is capable of showing repelling effects whose cause is DE (scalar field).

\begin{equation}\label{27}
\rho=\frac{1}{2}\dot{\phi}^2+V(\phi),
\end{equation}

\begin{equation}\label{28}
p=\frac{1}{2}\dot{\phi}^2-V(\phi),
\end{equation}
where the kinetic energy (KE) and potential energy (PE) of the scalar field are denoted by the expressions $ \frac{1}{2}\dot{\phi}^2 $ and $ V(\phi) $. As a result, the expression $\omega = \omega(t) $ \textit{i.e.} can no longer be used as a constant. With the observations given, $ \omega\simeq -1 $, the quintessence or phantom model agrees. As a result, we require $ \dot{\phi}^{2}<<V(\phi) $ \textit{i.e.}, where the $KE$ of $\phi$ is low relative to the $PE$. In this work, we assume that $\phi$ is the only source of DE with $V(\phi )$, therefore one can take into account energy density and pressure of scalar field as $\rho_{\phi}$ and $p_{{\phi}}$ respectively for flat FLRW space-time using Barrow's scheme \cite{jdbarrow}, employing Eqs. (\ref{27}) and (\ref{28}) as the sources
\begin{equation}\label{29}
\rho=\frac{1}{2}\dot{\phi}^2+V(\phi)=\rho_{\phi},
\end{equation}
\begin{equation}\label{30}
p=\frac{1}{2}\dot{\phi}^2-V(\phi)=p_{\phi}.
\end{equation}
By resolving Eqs. (\ref {29}) and (\ref {30}), the $KE$ and $PE$ can be found. The potential energy $ V(\phi) $ plots against the scalar field $\phi$ are shown in Fig. 4 for the same model parameter values as were used in Figs. 1 and 2. According to Fig. 4, the potential $ V(\phi) $ is present throughout the range $ -1<\phi<0 $ and $ V(\phi)\simeq0 $ at $ \phi\simeq0 $. As a result, we can assume that the only source of DE with a potential $ V(\phi) $ is the scalar field $\phi $. In light of this, we draw the conclusion that our model is an accelerating DE model. \\

In order to comprehend the nature of DE, we have assumed that the only source of energy is the scalar field $\phi$, which serves as the quintessence model. For this reason, we have examined our model in relation to the Swampland conjecture. The first conjecture of this criterion is related to the scalar field $\phi$, whereas the second hypothesis is related to the scalar potential $V$. The graphic in FIG. 9(a), 9(b) below show how our interaction model's scalar field behaviour behaves in relation to redshift. \\

\begin{figure}[H]
\begin{center}
$%
\begin{array}{c@{\hspace{.1in}}cc}
\includegraphics[width=3.5 in, height=2.5 in]{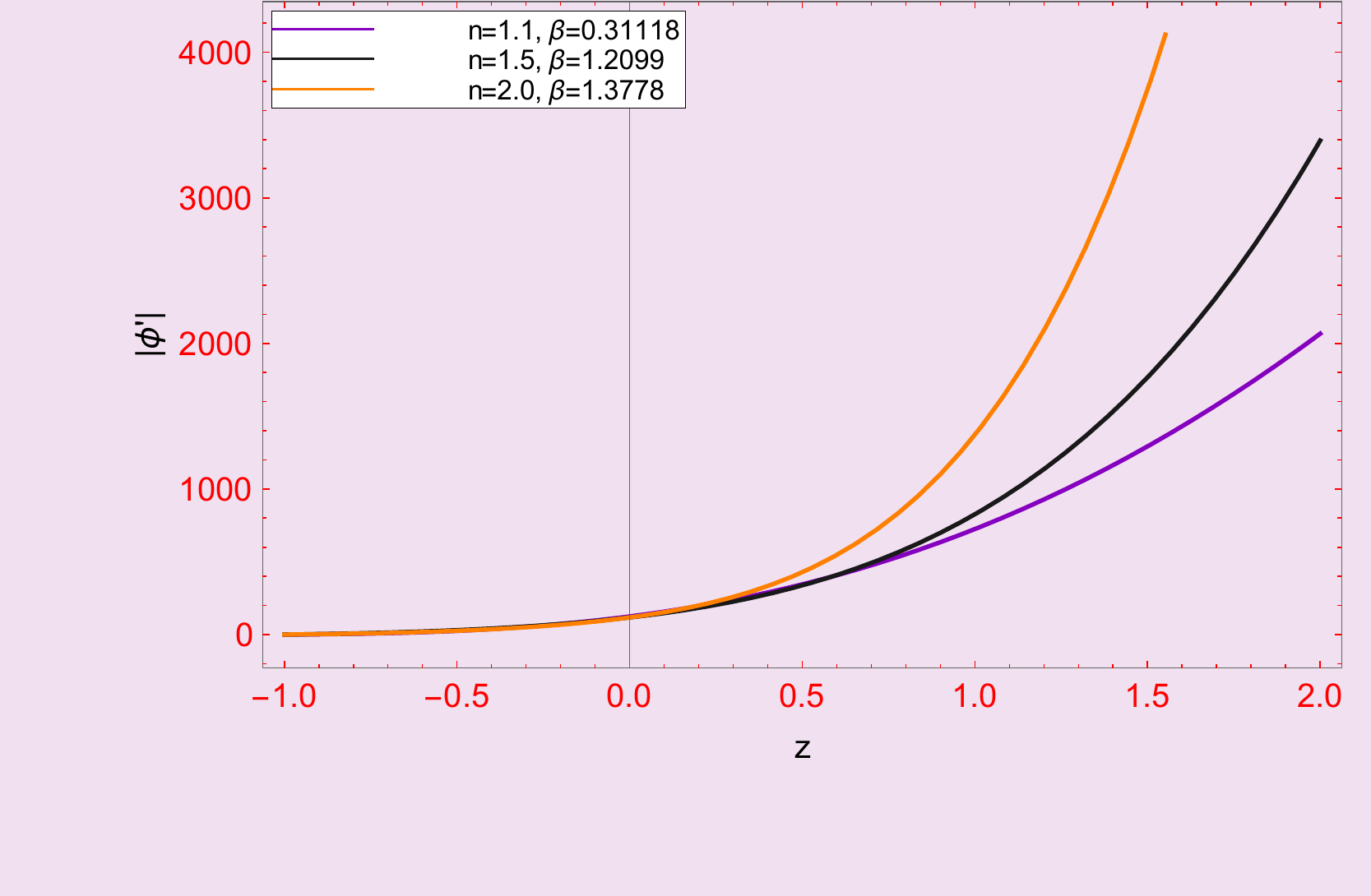} &%
\includegraphics[width=3.5 in, height=2.5 in]{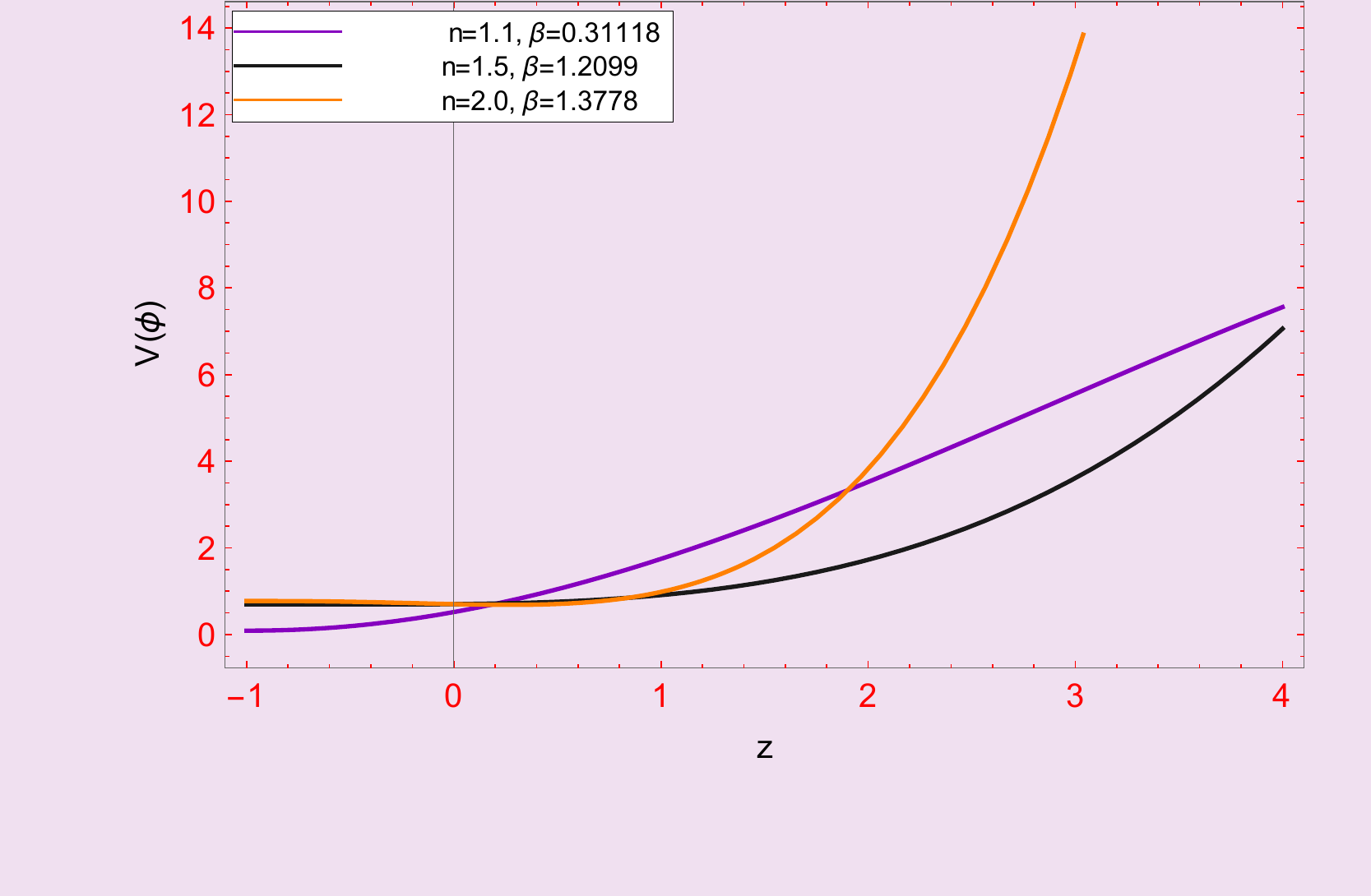} \\  
\mbox (a) & \mbox (b) %
\end{array}%
$%
\end{center}
\caption{\scriptsize The plots of $|\phi^{'}|$ and potential $V(\phi)$ w.r.t. $z$ for the model.}
\end{figure}
The image below in FIG. 10(a), 10(b) show how the components of the improved swampland conjecture behave when displayed for two different potentials linked to an interacting scenario for different cosmological parameters. We are aware that each of the theoretical swampland conjectures has positive unit order values and that component $C_2$ is smaller than component $C_1$. The figure illustrates how the interacting model shows that both hypotheses are satisfied. This DE model complies better with the Swampland criteria, as shown by the Swampland criterion, which has been given for a consistent theory of gravity in this regard and which meets the requirement $\frac{|V'|}{V }> C_1 \approx O(1)$.

\begin{figure}[H]
\begin{center}
$%
\begin{array}{c@{\hspace{.1in}}cc}
\includegraphics[width=3.5 in, height=2.5 in]{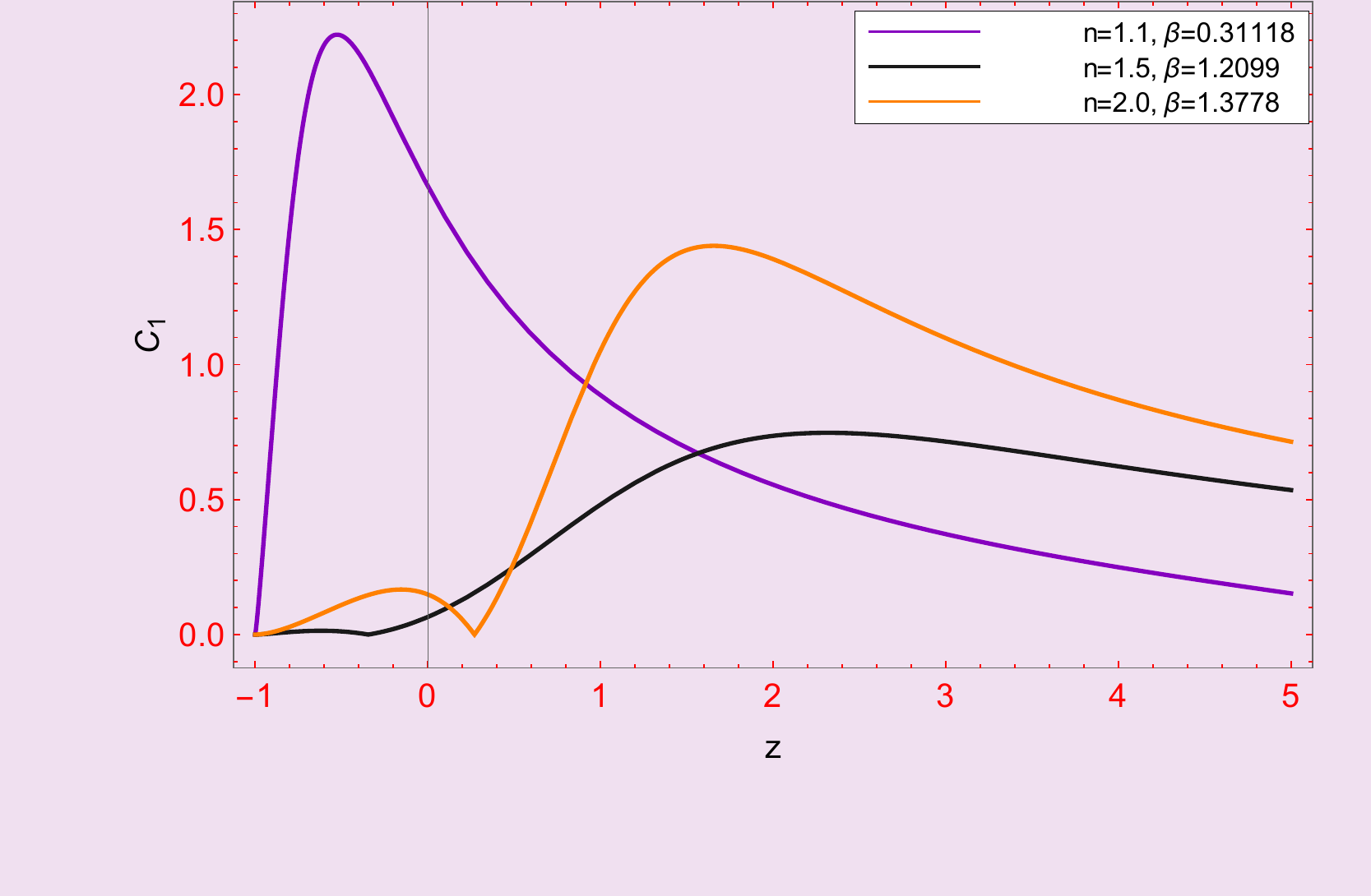} & %
\includegraphics[width=3.5 in, height=2.5 in]{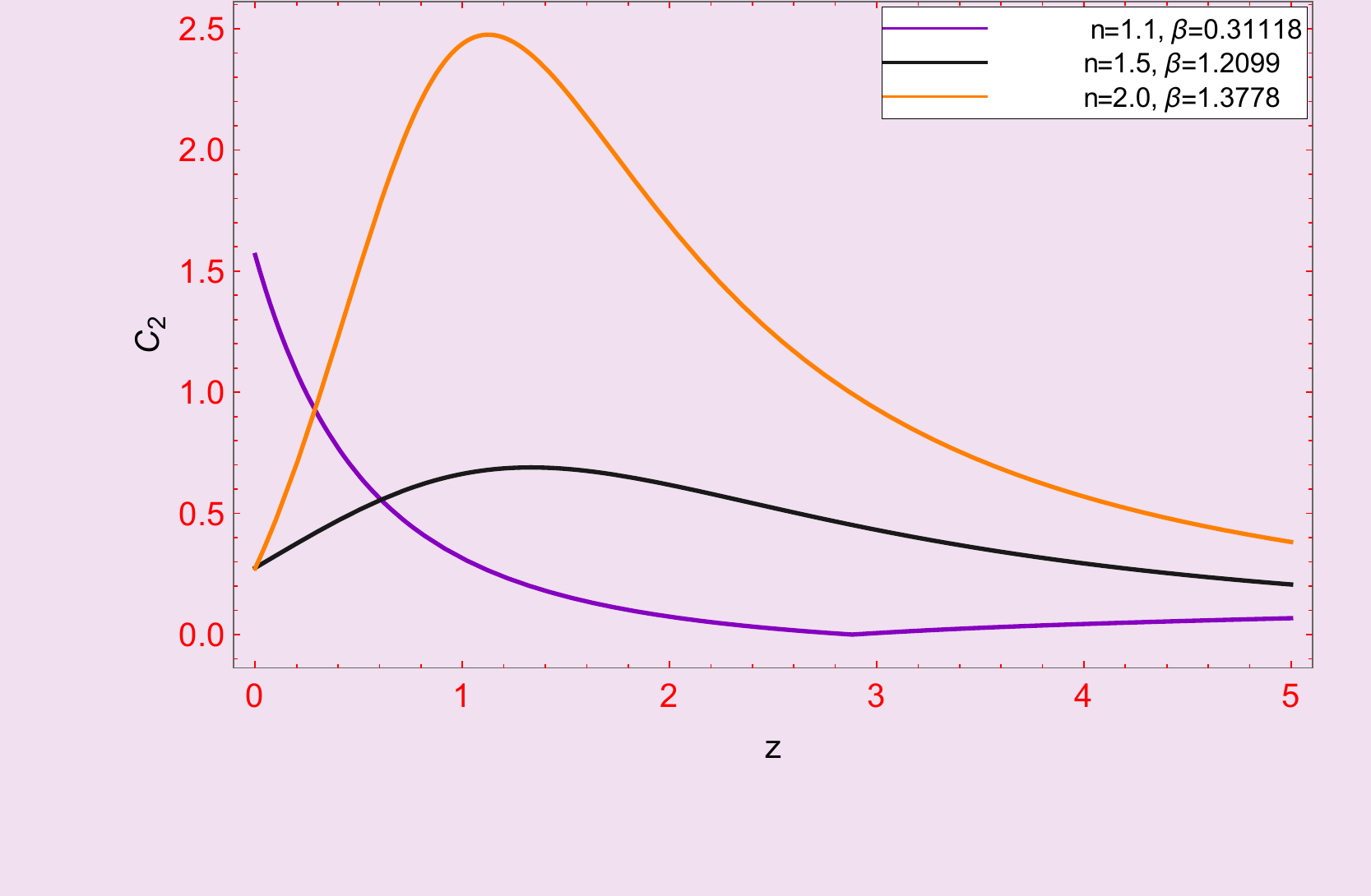} \\  
\mbox (a) & \mbox (b) %
\end{array}%
$%
\end{center}
\caption{\scriptsize The plots of swampland conjecture 1 and 2 for the model.}
\end{figure}

The above FIG. 10(a), 10(b) illustrate the behavior of each of the refined components of the swampland conjecture for different potentials linked with interacting scenarios for a flat Universe with different model parameters $n$. In the interacting scenario, we found the potential and created swampland hypotheses based on the EoS. We are aware that each component of the swampland hypotheses in the literature has positive values of unit order and that the second component $C_2$ is smaller than $C_1$. \\

According to the swampland conjecture, string theory should satisfy particular criteria in order to be considered a consistent quantum theory of gravity. According to the conjecture, coherent theories of quantum gravity must adhere to a number of constraints in order not to fall into the ``swampland" of inconsistent theories that cannot be effectively coupled to gravity. The swampland conjecture seeks to establish a direct connection between the potential energy density of the DM-DE interaction theory and its scalar field $\phi$. As we have already mentioned in the section VII that in order to understand the nature of DE, we have assumed that the $\phi$ is the only source of energy, serving as the quintessence model. For this reason, we have examined our model in relation to the Swampland conjecture. The first conjecture of this criterion is related to the scalar field $\phi$, whereas the second hypothesis is related to the scalar potential $V(\phi)$. The graphics show how our interacting model behaves for different values of the model parameters. The behavior of swampland conjectures for $n=1.5$ and $\beta=1.2099$ (the best fitted value of $n$ constrained in our previous work \cite{ritzannals}) is in favor as swampland conjecture meets the requirement $\frac{V'}{V} >C_{2} \approx  O(1)$ and that component $C_2$ is smaller than component $C_1$.


\section{CONCLUSIONS}

By assuming a specific type of cosmic evolution scenario, this article attempts to build a cosmological model. The merits of the parametric reconstruction method have previously been well discussed, therefore using this technique can help one to build a cosmological model that can predict probable phase changes and thus gain a better understanding of the underlying aspects of the cosmos. So in a way, the parametrization approach concept is pretty reasonable, and transparent, and helps in improving future efficiency. The motivation for considering the interacting scenario under the parametrization technique has been discussed in the introduction section. Accordingly, we have listed the key findings of our model analysis as follows:\newline

In the classical theory of gravity, we have looked at an interacting cosmological model of DE. This article provides an understanding of how cosmic parameters have evolved from the interaction scenario. The considered parametrization of the scale factor allows us to examine the late history of the Universe and give a seamless transition from early deceleration to present acceleration. We used several choices of $n$ (as found in our earlier work \cite{ritzannals} in this study to limit all the physical and geometrical parameters.\newline

We display the evolutionary patterns of trajectories of radiation, baryonic, DM, and DE density parameters for different $n$. As well as the energy density of different components of the Universe, the pressure of DE and EoS DE have been demonstrated. We can see that the radiation energy is dissipating more quickly than baryonic matter and DM, respectively. The profile of DE pressure, which is more negative in the past, rises concavely upward for $n=1.1,1.5$ and concave downwards for $n=2$, but finally, all three trajectories tend to zero in the future is highlighted in FIG. 2(b). \newline

Eventually, evolution will favor the conventional lore because of the dark energy's negative pressure. Figure 2(c) depicts the redshift history of the EoS DE, which is approaching $\approx -1$ in the far future for $n=1.5,2$ whereas the current value of $\omega_{DE}$ for $n=1.1$ is $\omega_{0,{DE}}= -0.472614$. This implies that $\omega_{0,{DE}}$ assumes the negative values in the anticipated ranges around $z = 0$. We have examined various energy requirements for interacting models and shown them with graphical representations. We have also addressed the sound velocity for our produced model using a reasonable selection of acceptable parameters, which is depicted in Figure 4. In addition, we addressed the improved swampland conjecture and recreated the scalar field's potential. A model of dark matter-dark energy interaction with cosmic consequences. In our study, swampland conjecture seeks to establish a direct connection between the potential energy density of the DM-DE interaction theory and its scalar field $\phi$.

\end{document}